**TITLE PAGE**

# Title: C-BPMN: A Context Aware BPMN for Modeling Complex Business Process

**Name of the Authors:** Debarpita Santra, Sankhayan Choudhury

**Affiliation and Address:** Department of Computer Science & Engineering, University of Calcutta, Kolkata, India

**Address:** University Campuses. Campus 8. Technology Campus. JD-2, Sector 3, Salt Lake City, Kolkata 700 098

**Email Address:** debarpita.cs@gmail.com

**ORCID of the author:** 0000-0001-5651-5642

## Abstract:

A complex business process demands adaptability as it has been highly influenced by the contextual information. The contextual information declares the underlying semantics on which the process logic depends. Thus one of the challenges of a business process modeling is to include the context sensitivity within the modeling itself. BPMN is the widely accepted tool in this field. All the process modeling languages like EPC, UML, BPMN are not able to express the context awareness as required. In this paper an attempt has been made to offer a means for modeling a complex business process with necessary contextual information. We have proposed a context model in terms of a graph, extended the existing BPMN by adding new construct and integrated the said components to achieve our goal. The methodology as stated certainly offers necessary understandability, maintainability and the adaptability as a whole. Moreover the model is validated using Colored Petri Net and is expected to behave properly in a real life environment.

## 1. Introduction

The Business Process Modeling Notation (BPMN) is an accepted standard to model business process flows and web services [1]. The specific objective of BPMN is to provide a notation that is readily understandable by all business users [2]. This BPMN diagram for a business process is designed to achieve two things. First, it should be easy to use and understandable by non-technical users. Second, it offers the expressiveness to model a complex business processes and can be mapped efficiently to business process execution language [3]. The organizational environment is truly dynamic in nature and it affects the undergoing business processes. The underlying dynamic behavior based on a specific environment may result divergence between the predefined process model and the current instances. Thus a business process needs to be adaptive to the varying context of the application scenario [4]. The need for making it adaptive demands the research on the evaluation on BP design through the inclusion of context sensitivity. The context has various meanings according to the application. Dey et al. [5] define context as “any information that can be used to characterize the situation of entities that are considered relevant to the interaction between a user and an application, including the user and the application themselves”. The process modeling languages like EPC, UML, BPMN are not able to express whether an activity in a business process is executed in a given context or not [6]. In these modeling languages, contextual variables are captured in text annotations or decision points that make the model unnecessarily complex [7]. As a result it becomes a challenge to offer a context sensitive process modeling notation to make a business process adaptive in true sense. The notion of embedding context awareness in BPMN has emerged as a pivotal research topic in Business Process Management over the last years. The interpretation of a process model will be incomplete without the context awareness as it fails to express the actual semantic within the business process. Understanding and analysis of the contextual environments in which business processes work needs a proper representation of the business process contexts. Thus the objective of this work is to include the context awareness within BPMN to make it more efficient for complex business process modeling.

The paper is organized as follows: Section 2 presents related works, Section 3 discusses the existing BPMN and its limitations, Section 4 describes the design of our proposed context model, Section 5 illustrates the design of context-aware BPMN model with examples, Section 6 briefly presents the simulation, implementation and the validation of the proposed model using Colored Petri Net, Section 7 makes some performance analysis of the model and Section 8 draws the conclusion.

## 2. Related Works

Researches on process adaptability against dynamically changing environments are being carried out since last fifteen years. The early researches were focusing on intrinsic context-related issues of a business process to increase its flexibility, the authors in [8] opine that the process flexibility should consist of an extrinsic trigger for change with necessary intrinsic adaptation mechanisms within the model. To increase the process flexibility and lower reaction time, the knowledge about the types of contexts and how the contexts could make changes in process level should be used and applied in early phases of the system design. To illustrate the notion of context-awareness in business process models, the authors stated that contexts of a business process can be prior anticipated and a flexible process adaptation should be done based on the change. For achieving greater flexibility, according to them, contexts should be captured externally from the business process model for ease of maintenance and ease of extension. Based on the research done in [8], the authors in [7] have proposed some extrinsic drivers for achieving business process flexibility along with an elaborate discussion on why and how context effects on the appropriate execution of a business process and how to incorporate contextual adaptability in the business processes. The authors opine that in most cases, the business process models are disconnected from their relevant contexts and it becomes difficult to track the situation in which the process takes place. In most modeling practices, relevant contextual variables become an integral part of control flow, leading to unnecessary structural and logical extension of the model which makes a mess with combining run-time and build-time decisions altogether. This kind of modeling practice reduces the acceptability of the process models by the end users when they do not want to be exposed to all the alternative decisions in the routine execution of the process. Another common practice is that, multiple process models are developed to cope with different contextual situations; but this kind of practice is inefficient as a huge redundancy is introduced by this modeling practice. The authors proposed a reference frame for the improvement and extension of business process modeling technique, so that context-awareness can be incorporated conceptually in the business process models. The proposed meta-model (reference frame) shows how the business process goals determine the set of relevant contextual factors so that the goals can be achieved in a well-defined manner. As per the meta-model (onion model), the contexts have been organized in four layers namely immediate, internal, external and environmental contexts, based on the characteristics of contexts. The onion model also takes into account some kind of inter-relationships among the contexts, where an element from the same or lower context layer can mediate, moderate or mitigate the impact of a context element. The authors also proposed a methodology on how different types of contexts can be incorporated in the business process models according to different goal-related information. However, this research was done in an explorative stage and the authors stated some future directions of the research. The researches [7, 8] have motivated us to propose an external context model which captures different kind of complex inter-relationships among the context elements and externally integrate the context model with the BPMN model to achieve greater flexibility in the model.

Some researches [9 - 11] are done on incorporating context-awareness in role-based business processes. The authors have proposed a business process context model and presented an approach on how different role assignments would be performed for the processes in a business process model based on different contexts. According to the authors, adaptability against the dynamically changing contextual situation requires frequent alteration in assignment relations according to situation. The authors introduced the concept of ‘context related knowledge’ with an aim to make the business processes more active, flexible and more expressive about using different types of business rules in different contextual situations. To incorporate context awareness among the business process models, an approach has been proposed where the instances of business processes are adapted to different changing situations and adaptation decisions (assignment activations) at each instantiations are based on context related knowledge. This kind of knowledge spans over six issues: ‘who’, ‘what’, ‘why’, ‘how’, ‘when’ and ‘where’ and the knowledge can be viewed from different aspects like temporal aspect, location aspect and so on. With this knowledge, a context model in terms of tree has been designed so that the relevant context facets, attributes and functions can be properly organized and measured. The context model uses first order logic to represent the contexts. At a given time, the original context model is adapted to hold only those facets and attributes of contexts that are relevant for a business process modeling at that time. Based on the adapted context tree, best role assignments are selected to instantiate the business process model. The researches [9 - 11] have only considered role based business process, but in real-life situations there exist many business processes which are activity or goal driven. To incorporate context-awareness in these business processes, the context model should be extended and redesigned, and there should be consideration of many adaptation strategies from different aspects to make these business processes more flexible.

The research [12] gives emphasis on the flexibility and adaptability issues of business processes whose execution evolves according to different enterprise requirements at different situations. The authors performed a survey on the existing approaches for context-awareness and found that using these approaches, it is difficult for each business process instances to support variability for different economic, technological or environmental contextual requirements. So, the context related knowledge is an essential resource to ensure adaptability among the business processes as a conventional business process model satisfies the customers' needs in a given context but not in other context. Integration of context-related knowledge with a business process model allows the model to be active, flexible and fine-grained. For the survey purpose, the authors considered first those researches in business process modeling where role is the main concept for representation of whole model. But, the authors opine that the approaches [9] that only deal with role descriptions, are not satisfactory to meet the flexibility requirements compared to the approaches [13,14] that represent roles as set of ordered activities or interactions. The authors come to conclusion that it would be useful to adopt role based methods in business process modeling if these roles pose sufficient flexibility to meet the organizational, functional or operational requirements. The second set of models for the survey is goal-oriented. Goals pose high-level objectives of an organization and these goals are specified in a model in terms of actors and activities. The survey also includes activity oriented model which is basically a set of activities along with their relationships regarding the pre-defined control and data flows. The formalism is mainly useful for representing the functional view of business processes. Despite its inadequacy for modeling ill-defined business processes that are exposed to frequent changes, the activity-oriented models are still dominant in the literature. The authors stated that flexibility can be achieved on the modeling formalisms by two ways: at design time (a priori flexibility) and at run time (a posteriori flexibility). When a model is able to cater with environmental changes without any evolution of process definitions, then the model is supposed to be incorporated with this capacity inside the process definitions at design or build time. Run time flexibility in a process model is required against changes that impact the process definition or instances. In this case, addressing these changes in the process model requires either a dynamic adaptation performed on one or several instances when the process definition is not convenient for the execution conditions or a correction on the process definition of an exception which happens during the execution of an instance. Sometimes, an evolutionary transformation is needed due to the redesign or reconfiguration of the business process. Versioning technique is also there if a modeling approach has to handle several versions of the same process definition as an adaptation mechanism for run-time flexibility issues. Versioning supports the concept on business process evolution and anticipates future transformations.

The authors in [15] started their research with an assumption that the business process model is already designed and there is a need to evaluate and re-engineer the business process model under constantly changing contexts. To address the need, the authors extended the existing redesign mechanism by associating to each workflow pattern of the model a degree of relevance according to particular contexts. The reason behind taking into account the work flow patterns (control, resource and data patterns) is that they provide a formal basis for understanding the control flow requirements in the business process model and for evaluating the capabilities of business process modeling languages. The authors proposed an evaluation framework for work-flow patterns under different contexts, where a context captures and determines a process nature (production, administrative, collaborative and ad-hoc) and the degree of relevance for each workflow pattern vary from 'Not at all important' to 'Very important', with two more degrees in-between. The authors evaluate the business process instances to discover various features of a process context. This work of identifying weakness of a business process model and redesigning the process model according to adequate selection of workflow patterns with respect to different process contexts is one of the early researches recorded in the literature for context-aware business process re-engineering. This research has significant differences from our proposal as our consideration of contexts is not only confined into the process level, but also it is expanded in many broader aspects like organization, role and external factors. Also, adaptations beyond the workflow changes are required in a business process model to cope with dynamically changing situations.

While investigating how context-awareness can become an integral part of business process modeling [4,7,16], researchers in [6] proposed an approach to know in better way the variables that play the key role for contextual adaptability in the business processes. These variables are utilized for improving performance of a business process model in terms of cost, time and level of service. With the knowledge of contextual variables and about their potential impacts on business processes, more detailed analysis can be done on how a process can be adaptable to continuously changing environment. The authors analyzed how previous incidents can be learnt by the business process model so that it can show better adaptability when similar incidents occur. The authors also discussed about how a business process model can automatically learn to improve its behaviour in changing environments and accomplish the goal without fail. But learning is a complex process which involves lots of time and huge cost. Learning is a repetitive process and many authentic training examples are required. So, when the time and cost are important issues while developing the model, this approach may not be a suitable option.

The research in [17] opts to provide a solution to any developed software to cope with variability in different organizational environments and hence the adaptability issues of the software. Variability differentiates between the common and different parts in a set of similar but different product lines of a product family. With the aim to managing the commonalities and variability among processes to reuse common parts and making the products adapted to different customers and different organizational settings, the authors proposed a representation system called 'MAP' to capture the variability across business processes of a family in an intentional manner. A business process family is basically a collection of processes meeting a common goal but in different ways. The proposed MAP has been represented as a directed, labeled and non-deterministic graph where nodes hold the goals of business processes and the multi-edges between the nodes are to bind the variability between the processes. So, there are many traversal possible in the graph between the start and the end nodes. The graph can also be viewed as hierarchically to capture different levels of the variability. Variability in business process family arises from the fact that there exist different strategies or ways of a business process to reach a particular business goal. This kind of variability comes from an intentional view of a process. Using the concept of feature which is a logical unit of behaviour of a business process, the MAP model basically offers the process capacity to deal with the environmental change without any evolution of process definitions at build-time and allows run-time adaptation in the concerned business process. The authors also proposed two kinds of adaptation strategies for particular combinations of features inside a MAP namely Design time adaptation and Run time adaptation strategies. Design time adaptation allows selection of a combination of features resulting in only one path from Start to Stop, and Run time adaptation addresses a large degree of variability in the adapted MAP and desired features can be selected dynamically at the beginning on the process. The authors think that expressing the variability with the MAP formalism is particularly useful at the adaptation phase, as the business process owner is exposed to the choices that are relevant to the satisfaction of his/her goals in terms of the properties of the business, with no need to deal with the details from organizational and technical configuration perspectives. This research indeed gives us a right direction; but it deals with a business process family, where process variability is of great concern. But this approach may not be applicable for an individual business process.

In [18], the authors proposed a framework, with the introduction of an artifact, to enable a context-aware design approach and applied the model in the organization environment of an Australian insurance provider. The framework would support process managers in making process changes with respect to the changing parameters. The authors were motivated by the lack of concrete artifacts to support context-awareness in processes that need to be adapted to changing contexts. Using the artifact, the authors want the process manager to analyze the feedback structure of a business process and its environment and identify suitable adaptation strategies for context-awareness. The artifact would basically capture the process of coupling between the external and internal process variables and the impact of changes on internal variables due to external variables. As a business process can be thought of as a non-linear process, the solution stated in this paper considers both the high-level feedback structure of the system as well as individual actions inside a business process. The authors first model the 'macro-level' viewpoint of the system which basically models the system variables and their interactions. After that, they model the 'meso-level' viewpoint of the system which basically defines the global observation behavior of the system. Finally, both of these specifications are assembled into the overall model using feedback structures with an aim to analyze how process activities in a business process are affected by system variables and vice versa. This research is one of the early attempts to incorporate context awareness into a business process by using feedback loops between the external and internal process variables. The proposed approach seemed suitable for the Australian Insurance provider company but needs to be generalized so that the approach can be applied in versatile domain.

Motivated by the explicit considerations of external variables in process design [7], the research paper [19] offers an extended conceptualization of business processes as complex adaptive systems with the aim to optimize the processes by analyzing the process operations in different contexts and by examining the complex interaction among the external contextual elements and the internal process schema. On a practical level, this study provides an early insights into the challenges that the organizations are facing to cope with dynamic process environments. For this research work, the authors have chosen core processes of two different organizations like an insurer and an airport system and analyzed how these processes can be adapted at regular intervals to frequent externally induced changes like severe weather events or cycles in domestic and global economic activity. In both the cases, the processes interact with a dynamic and changing requirement and the challenge is to adapt their behavior to offer the same quality of service in different contexts. According to the authors, when a process is conceptualized as open, adaptive complex system, it requires addressing three challenges: a) discovering and understanding context, b) analyzing the impact of context change and c) enforcing corrective action. After careful observation of both the case studies, the authors claim that contemporary configuration mechanisms are not sufficient to combat the constantly changing process requirements in a dynamic environment. The authors suggest that the current body of process knowledge should be extended by context-aware artefacts to visualize the full traceability of context changes and their impact on the activities and resources of the process and to enhance the process adaptive-ness by linking context changes to configuration parameters of the process model. The authors also suggest that context variables should be included into the business process management life cycle. The context variables should be included in the analysis, design and implementation stages of the lifecycle to support the need of adaptive-ness and defining suitable adaptation strategies inside the process.

In [20], the authors have conceptualized the structure of a business process as a set of rules in the form of 'Event – Condition – Action' (ECA). The business rules are derived based on dynamic changes in the business context. Using these rules, the authors proposed a reliable method for executing dynamic business process adaptation, according to eventual modifications in the context information. The authors stated that according to context, the context aware system dynamically adds to the business process instance some activities. To allow this modification, the ECA rules for these transitions are modified. But the paper lacks clarity about the methodology followed for the modification at run-time. Also, the CFC table stores lots of null values, making the whole approach inefficient to some extent. Each time a new context occurs in the business process, a new CFC table with long size is generated and also the comparison for each entry in both the old and new is done. The comparison process incurs a lot of computational time, making the process computationally inefficient. Also only the users' contexts are considered in this model. There should be many other contexts to be taken into consideration and different contextual requirements require different kind of adaptations.

In [21], the authors proposed a meta-model for formally specifying functional requirements for content and context-aware dynamic service selection in business process models to achieve greater flexibility. The dynamic service selection for a business process model is necessary to cope with the dynamically changing contexts. The motivation behind this research is that development of a business process is affected by its dynamic nature, which indicates that dynamic binding is necessary between the model and selected services according to contextual changes. The authors claim that business processes that would leverage the proposed meta-model can be able to changing circumstances without any need to make changes in the process flows. The proposed meta-model has been implemented using BPMN 2.0 and allows users to configure it in run-time. But dynamic selection of services and their composition for each time may be a costly approach for some of the business process models.

The authors in [22] have proposed an integration methodology for context constraints with process related role- based access control (RBAC) models to incorporate the features of context-dependent task execution inside the models. For this research work, the authors have defined some process related context constraints and incorporated the contextual notions in the extended UML based activity diagrams. But, no adaptation polies according to the contextual changes are specified in the paper. Also the authors have considered role-based contexts and how each activity is affected is not mentioned.

In [23], the authors proposed a comprehensive framework which uses context-aware composition of process fragments to develop an adaptable service-based application like business process model. The framework promises and offers a set of adaptation mechanisms to solve complex adaptation problems of real life. The framework suggests that a business process is partially defined at design time in terms of abstract activities for achieving the goal and is automatically defined at run-time based on the contextual situations. During this refinement, available fragments for the specific contexts are supplied by different providers. The adaptation mechanisms that are specified in this research work are based on some extensions of classy AI planning techniques used for automated service composition [24]. The authors use a shared context model which describe the operational environment of the system and, each of the context is defined through a set of context properties. The authors have introduced a term 'context configuration' to denote the present status of all the context properties of a context at a specific time. The proposed adaptation model uses the idea that different entities are added in the business process dynamically and the functionalities of the entities are published through the dynamically selected process fragments. In the business process model, each activity is annotated with 'preconditions' and 'effects', where the preconditions impose constraints on the execution of an activity under specific context configuration at run-time and, effects model the expected impact on activity execution. But use of 'abstract activity' is not a good idea always, as this needs to be refined always even when there is no requirement of executional modification of the underlying activity. This is a costly and time-consuming process. Also, every time composition among the process fragments are done, which is also a time consuming and costly process.

Another attempt for incorporating context awareness in business process modeling in the paper [25]. The authors have proposed a hierarchical ontology for general business process context along with proposal of a context-aware BPM framework. But the adaptation mechanisms are not clearly mentioned in the paper. Despite many such approaches, improving business processes with the help of application of context-awareness is an ever-demanding requirement to offer a smart and intelligent environment to the society [26]. In [27] authors have presented a thorough overview on the issues of consistency that exist among multiple variants or instances of the same business process model. The authors have analyzed the key factors behind inter-model consistency and generates some research questions. The authors also considered the elicitation of consistency requirements in context-aware business processes, where the source of inconsistency is the context related knowledge. The authors also proposed a context-aware road map which would support consistency requirements elicitation and management for business process modeling. This road map is expected to guide building the right system without any inconsistencies. The authors in [28] have proposed a prototype named ArchReco, which is basically an educational tool to recommend employing context-awareness in design patterns required for software development. Though this kind of context consideration is different from ours, all the researches that are being carried out, are for offering more flexibility and adaptive-ness in the business process models and the research would go on.

In this paper, we assume that a business process model is designed for ideal situation. The business process designers define at design time a particular flow of executions for a particular environment. This environment offers an ideal situation. When a change is sensed in the environment violating the ideal situation, some of the pre-defined set are activities cannot be of no use and therefore, adaptations of the activities are needed with the changing scenario. These activities can be replaced by another set of activities, can be omitted or extra activities can be added, with all other activities and other business process elements remaining intact. In our approach, a business process context model is designed and an extension in BPMN is proposed for attaching the context model externally with the extended BPMN model.

## 3. BPMN Model and its Limitations

An organization models a business process to achieve a typical business goal. For the modeling, there should be clear specification about activities, their execution sequence as well as the execution constraints, types of resources needed and so on. A business process modeling requires standardized notation so that the model can be readily understandable by all business users. The Business Process Modeling Notation (BPMN) is a well-known standard to model business process flows and web services [29]. This includes the business analysts that create the initial drafts of the processes to the technical developers responsible for implementing the technology that will perform those processes. This BPMN diagram for a business process is designed to do two things well. First, it is easy to use and understand the diagram by non-technical users. Second, it offers the expressiveness to model very complex business processes, and can be naturally mapped to business process execution languages.

To model a business process using BPMN, one can simply use the events that occur to start an initial process, to trigger the processes that get performed in the intermediate, and the end results of the process flow. An event either kicks off a process flow, or happens during a process flow, or ends a process flow. A process in BPMN is also called an activity. Business decisions and branching of flows is modeled using gateways. A gateway is similar to a decision symbol in a flowchart. Furthermore, a process in the flow can contain sub-processes, which can be graphically shown by another business process diagram connected via a hyperlink to a process symbol. If a process is not decomposed by sub-processes, it is considered a task which is called the lowest-level process. A '+' mark in the process symbol denotes that the process is decomposed into multiple sub-processes or sub-tasks. If it doesn't have a '+' mark, it is a task. Driving further into business analysis, one can specify 'who does what' by placing the events and processes into shaded areas called pools that denote who is performing a process. One can further partition a pool into lanes. A pool normally signifies an organization and a lane usually characterizes a department within that organization [30].

BPMN provides many constructs so that a business process can be represented diagrammatically in efficient way. There are distinct notations in BPMN for various types of events, gateways, data objects to model a business process. BPMN also allows us with a textual annotation that can be affixed to any model element, so that one may describe extra details about the element in good old-fashioned words. So, in one word, we can claim that BPMN is designed to enable modelers to easily model typical business processes and offers the capability to model complex business processes, including the message passing of web services [29].

Despite the advantages of BPMN, the notion of context awareness in BPMN has emerged as a pivotal research topic in Business Process Management over the last years. The need for augmented consideration to flexibility to cope with dynamically changing situation originates from two major contributors. First, the tendency to declining time-to-market and time-to-customer demands and a growing occurrence of product modernizations associated with market changes such as globalization and new stages of compliance necessitate adaptive business processes [31]. In simple terms, flexibility provides the ability to change without loss of identity [32]. Business process flexibility can be seen as the capability of a process to respond to externally activated change by revising only those aspects of a process that need to be modified and keeping other parts stable, i.e. the skill to alter the process without totally substituting it [31]. Current BPMN modeling technique does not consider contextualization. It ignores the spur for change occurred from inside or outside of the organizations. But this motivation for change should be considered. The motivation for context-consideration in a business process model is that it leads to a sturdier cause-effect relationship between the need for process flexibility and their effects on processes and vice versa. Relevant changes in the business environment can be predicted and subsequently activate the timely adaptation of business procedures. Hence, explicit and clear context awareness inspires observing of the relevant process context. The initial identification of context changes together with knowledge about what type of process changes are essential points to more process flexibility requirements as well as declined reaction time and upgraded risk management [11].

Throughout the paper, we will consider a running example of a kiosk-based remote healthcare scenario in India. From the broader business aspect, the processes or activities within a health kiosk form a sequence or chain to achieve the goal. At first, patient comes at kiosk and a receptionist registers the patient with his (her) basic information like name, age, gender etc. After the registration (manual), a healthcare assistant records the symptoms along with other relevant information about a patient. Based on the patient's medical information, a caregiver starts the treatment process of the patient. After a patient is diagnosed, all the records of the patient (registration information, medical information and outcomes of the treatment) are uploaded in cloud server using the available network by another healthcare assistant. After a patient is diagnosed, the patient pays the bill for treatment at the kiosk. This scenario has been modeled using BPMN as shown in figure 1. In this paper, we are considering an activity-driven business process model. Here, we use the term 'Activity' and 'Process' interchangeably. This BPMN model shows basically the flow of activities. For the sake of clear visibility, we have not used the Swim-Lane constructs of BPMN. Instead, we have color coded the activities based on different colors of different roles.

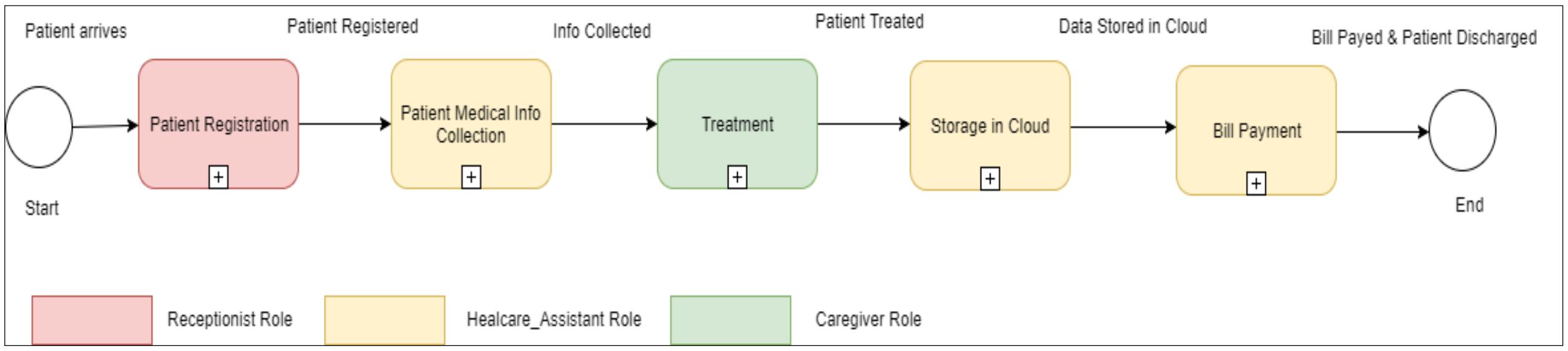

Figure 1: Simple BPMN model of a Kiosk-based Healthcare Scenario in India

## 4. Proposed Business Process Context Model

### 4.1. What is a Business Process Context?

A business process context is defined as the information that impact the design and execution of a business process either externally or internally. This can be regarded as a specialization of the definition given by Dey [5]. A business process context may be static, steady or dynamic depending on application.

**Static Context**: The context information whose value is fixed for all the time is called static context. This kind of context information does not change at all with respect to time. For example, in our application scenario, the name and date of birth of patients do not change. So, 'name' and 'date of birth' are static contexts.

**Steady Context**: The context information whose value becomes changed after a certain time interval (not frequent) is called steady context. The duration of the time interval will be application specific. This kind of context information remains static or fixed for some time and changes occasionally. Examples of Steady Contexts are: a) age of a person which changes every year, b) no. of employees in a healthcare kiosk, which changes in six months when new persons are recruited, c) no. of available instruments there which also may change occasionally after some new ones are bought.

**Dynamic Context**: The context information whose value is changed frequently e.g. within a very short interval from one minute to the next, is called dynamic context. In our example scenario, 'weather', and 'network status' are two examples of dynamic contexts.

There may exist dependencies among contexts. This kind of dependency sometimes trigger another contextual change in the environment. For our example scenario, we assume that internet connection remains available most of the time in the kiosk when the weather is good (i.e. no rain). But when it rains, the connection becomes unavailable. So, if 'network_status' is a dynamic context to represent the internet connection status and 'weather_status' is another dynamic context to represent the weather condition, the former context is said to be dependent on the latter context. A context may be dependent on more than one contexts. Also, new contexts can be derived from the existing contexts according to some pre-defined rules. For example, if the patient registration process is completed and patient is now in the treatment room in the presence of care-giver, it can be easy to infer that the treatment of the patient is going on. So, the value of the context 'patient_treatment_status' will be '*running'*, only if the value of the context 'patient_registration_status' is '*completed*', value of the context 'patient_location' is '*treatment room*' and the value of the context 'caregiver_status' is '*present*'.

### 4.2 Categorization of Business Process Contexts

As mentioned earlier, a context basically reflects the changing circumstances during the execution of a business process. Many research have been done on identifying various kind of context factors that affect a business process model. Rosenmann.et. al. [6] categorized the contexts into four layers: Immediate contexts, Internal Contexts, External Contexts and Environmental contexts. The immediate context includes those elements that constitute the pure control flow, and covers those elements that directly facilitate the execution of a process. The internal context covers information on the internal environment of an organization that impacts the business process. The external context captures elements that are part of an even wider system whose design and behavior is beyond the control sphere of an organization, and the environmental context which resides beyond the business network in which the organization is embedded but yet effects the business processes. [33] divided business contexts into eight categories: i) context of the business process: business activity, interaction among business entities, ii) context related to products, iii) context related to markets, iv) the geopolitical contexts, v) context related to legal or govt. constraints, vi) context related to roles, vii) context related to support roles and, viii) context related to system capacity. Born et al. [31] give examples of categories, e.g. industry, geography, period but did not confine into fixed set of categories. Saidini et al. [9] proposes four types of context information: i) the context related to location, ii) context related to time, iii) context related to resources and, iv) context related to organization.

Considering all the above-mentioned contexts and keeping in mind the ease of context representation for an activity-driven business process model, we have categorized the contexts that affect the flow of activities in a business process model from three different perspectives: Context related to Organization, Context related to Role, Context related to External Factors. The contexts from organizational perspective can be classified into structural contexts, goal related contexts, resource related contexts, data related contexts and product related contexts. Contexts from role's perspective include both the actors and non-actors (partners, competitors etc.) using or participating in the business process. Contexts from external factors' perspectives are place, temporal, season, geopolitical and many other social and environmental factors. Context information show a number of temporal characteristics. While the values of static contexts can be directly obtained from the concerned users, values of steady contexts are derived from the values of static or dynamic contexts. Dynamic context values can be obtained through sensors or other sources. There exists context provider who provides the contextual values at different time instants.

**4.3 Proposed Business Process Context Model**

A context ***C*** can be formally represented by 4-tuple $\langle p, a, l, v\rangle$, where $p$ is a context parameter, $a$ is an attribute of the parameter, $v$ is the value of the corresponding attribute and $l$ is the connector between the attribute and the corresponding value. The connector can be any comparison operator like =, >, <, ≥, ≤, ≠ etc., any preposition in English like ‘In’, ‘At’ etc., or any adverb like ‘near’ etc. [42]. Example of a context is <Weather, Temperature, >, 30$^o$ C>. This kind of context is called ‘Atomic Context’. We can obtain a ‘Composite Context’ by combining the atomic contexts using the Boolean operations AND and OR.

A business process model becomes affected by many atomic contexts at any instant of time. The effecting contexts of the business process model at a particular time instant is a vector CON($t_i$) = [$C_1$, $C_2$,…,$C_n$], where $t_i$ is any time instant and $n$ (> 0) is the total no. of effecting contexts and $C_i$ is an atomic context with $1 \le i \le n$. Suppose, $t_j$ is the current time instant and CON($t_j$) holds the effecting contexts on the business process model at time $t_j$. A ‘Contextual Situation’ (**CS**) at time instant $t_j$ holds basically the change occurred in the CON($t_j$) with respect to the previous context vector CON($t_i$). Here, we mention that $i$ may or may not be equal to ($j$ -1) depending on the dynamically changing situation of application-specific business process models. The idea behind introducing the concept of ‘Contextual Situation’ is that we need to identify who (context parameters, attributes) are responsible for two consecutive context changes. If we observe that at time instant $t_k$ ($k$ > 0), the contexts effecting on a particular business process model is CON($t_k$) = [$C_a$, $C_b$, $C_c$], where $1 \le a, b, c \le n$ and $a \ne b \ne c$, and at time instant $t_{k+1}$ the effecting contexts are also $C_a$, $C_b$ and $C_c$ with CON($t_{k+1}$) = [$C_a$, $C_b$, $C_c$], we would say that there is no context change observed at time $t_{k+1}$. So, CON($t_{k+1}$) = CON($t_k$)**.** Suppose at next time instant $t_{k+2}$ we observe CON($t_{k+2}$) = [$C_a$, $C_b$, $C_d$] ($1 \le d \le n$, where $d \ne$ a / b / c). We see that there is a change in the vector CON($t_{k+2}$) w.r.t. CON($t_k$). Contextual Situation at time instant $t_{k+2}$ holds this change. A change captured in the **CS** can be due to addition of new context parameters where some of context parameters from the previously effecting contexts may not be present, addition of new attributes under a same previous context parameter or change in values of attributes that previously affected the business process and also exist in the list of effecting contexts at the latest time instant. A Contextual Situation (**CS**) that can impact a whole business process, lists those newly added parameters, newly added attributes and the attributes with changed values.

A **CS** is represented mathematically as 3-tuple $<P_c, A_c, t_c>$, where $P_c$ represents a set of changed context parameters at time instant $t_c$ and $A_c$ represents a set of changed attributes (both the attributes that are newly added and the existing attributes whose values are modified) at the same time instant. $P_c = \{p_i, p_j,\ldots p_m\}$, where $p_{k'}$ ($k' > 0$) is a context parameter that affects a particular business process. It is an obvious and trivial assumption that a business process would be affected by only those contextual changes that are within the goal-scope of the process. More specifically, a business process that handles bank loan related services will never be affected by emergency situation of a patient. But this context would affect another business process that takes care of healthcare services. A particular business process administrator obviously has sufficient knowledge at pre-design time about the possible context parameters and attributes that would affect a business process. The second tuple in **CS** is $A_c = \{p_a.a_1, p_a.a_k, p_c.a_3, p_d.a_j \ldots p_f.a_l\}$, where an element of the form $p_g.a_h$ represents an attribute $a_h$ attached to a context parameter $p_g$, with $p_g \in \{p_a, p_c, \ldots p_f\}$ and does not belong to $\{p_i, p_j, \ldots p_m\}$. The set of context parameters $\{p_a, p_c, \ldots p_f\}$ were present in the CS at previous time instant and only those attributes whose values are changed at time instant $t_c$ are added to the list $A_c$. There is no need to explicitly include in $A_c$ the attributes associated with the context parameters present in $P_c$.

We introduce another concept called 'Context State' (**S**) for an individual activity in a specific business process. **S** is basically a sub-set of **CS** containing only those elements that affect only the particular business process activity. It must be noted that a business process activity is affected by a certain set of contexts, not all the contexts affecting the whole environment of the business process at a certain time instant. As mentioned earlier, the business administrator has the domain knowledge about the possible contexts that may affect individual activities in a business process.

We illustrate both the concepts of **CS** and **S** using simple example from our application scenario. Let us consider that a context change has occurred when it started to rain at 11.00 am and the weather was sunny just a moment ago. Suppose the last effecting contexts <Weather, Status, =, Sunny> <Watch, Time, =, 10.30 am> and, <Healthcare_Employee, Status, =, Present> were noted at time 10.30 am. At time instant 11:00 am, the effecting contexts are <Weather, Status, =, Rainy> and <Watch, Time, =, 11.00 am>. So, the contextual change is captured in the CS as [<Weather, Watch> <Weather.Status, Watch.Time> <Rainy, 11.00 am>]. This contextual change will not impact the activity 'Patient Registration', but it will have impact on the activity 'Store Data in Cloud' as the contextual change will trigger another context change as the status of network becomes unavailable due to rain. As a result, for the activity 'Patient Registration', no change will be made in the associated Context State $\mathbf{S_1}$ w.r.t previous time stamp. But changes will be made in the Context State $\mathbf{S_4}$ of the activity 'Store Data in Cloud' w.r.t. previous time stamp.

We now propose a graph-based context model that represents the nature of possible contexts and their interrelationship for a business process application environment. In this paper, we are concerned about the changes in a business environments w.r.t. time; hence we are considering only the steady and dynamic contexts. These contexts span all the contexts from organization, role and external factors perspectives. Capturing the behaviour and dynamicity of a business process's overall contextual situation is a complex task which leads us to develop the context model in 3-level hierarchies where the lower level is called the *Plane of Context States,* intermediate level is called the *Plane of Entity-Attribute-Relationship* and the upper level is the *Plane of Observation*. The *Plane of Context States* includes the context states corresponding to activities in a business process model. The *Plane of Entity-Attribute-Relationship* contains different types of context parameters along with their attributes. Relationships exist among the context parameters and among the attributes. The *Plane of Observation* is for reflecting the values of the atomic and composite context attributes.

The contexts present in our proposed context model are highly interrelated. In our example scenario, a common type of relationship exists involving a healthcare employee, the laptop she possesses and the internet connectivity that the laptop uses. As told earlier, another type of relationship is called the *Dependency Relationship* among the contexts, where a context or a set of contexts affect another context and the value of the latter is derived from the value(s) of the former. In this case, the latter context may or may not hold a value directly inputted by the context provider. As an example, we can think of a context 'Network.Status' which may hold the value 'available' given by the context provider. But, when a context 'Weather.Status' holds a value 'heavy rain', it will force the value of the context 'Network.Status' to be changed into 'not_available'. This kind of *Dependency Relationship* is termed as *Partial Dependency Relationship*. Here the latter context is said to be partially dependent on the former context. Consider another example where a patient can be admitted to a hospital only if patient beds are available. So, the value of the context 'Hospital_Bed.Availability' will entirely determine the value of the context 'Patient_Admission.Status'. The latter context which cannot accept any direct inputted value from the context provider, would hold the value 'admitted' only if the value of the former context is 'yes'. This kind of dependency relationship is called the *Total Dependency Relationship*. In this case, the latter context is totally dependent on the former one.

We define four different types of nodes in our proposed context model: i) *State* nodes (represented by bold circles), ii) *Entity* nodes (represented by rectangular boxes), iii) *Attribute* nodes (represented by small circles) and, iv) *Value* nodes (represented by ovals). The *State* nodes reside at the *Plane of Context States* (level 1) and each of these nodes represents a context state. The *Entity* and *Attribute* nodes reside at the intermediate plane (level 2), where an *Entity* node corresponds to a context parameter and the *Attribute* nodes correspond to the attributes associated with the context parameter. Among the *Entity* and *Attribute* nodes, some nodes fulfil the organizational perspective, some fulfil the role's perspective and rest fulfil the external factors' perspective. Two types of *Attribute* nodes are there. The attributes which represent dynamic contexts are represented by singly outlined circles and the steady contexts are shown by doubly outlined circles. When the *Attribute* nodes take values directly from the context provider, the outlines of circles become straight; otherwise the outlines become dotted. The *Value* nodes reside at the *Plane of Observation* (level 3) and consist of two types of nodes: one node for holding atomic context value (represented by thin oval) corresponding to each *Attribute* node present in the intermediate layer and another type of value node for holding the value of composite attributes (represented by thick or bold oval). The number of composite value nodes at level 3 will be same as the number of context states at level 1. The values of the context parameters and attributes enclosed in the context states will be observed from the value nodes. Each composite value node would hold the values of contexts corresponding to a context state in level 1.

At level 2, relationship between entities is represented by straight lines and the cardinality relation among the entities through the relationship can be one-to-one, one-to-many, many-to-one or many-to-many. The straight line may be directed or undirected depending on the mapping constraints of entities. In case of one-to-many and many-to-one constraints, the 'one' sides have the direction. In case of one-to-one constraint, every side of the straight line is directed and in case of many-to-many constraint, both sides of the line are undirected. The partial and total dependency relationships among the attributes are represented by directed dotted curve and directed straight lined curve respectively with both being annotated with corresponding arc expressions. The three levels in the model are connected by directed links originated from the lower hierarchy to upper hierarchy. Between level 1 and level 2, there exists two kinds of links: red colored arc for mapping between a context parameter enclosed in a context state (level 1) to an *Entity* node (level 2) and blue colored arc for mapping between an attribute inside a context state (level 1) and an *Attribute* node (level 2). Between level 2 and level 3, there exists another kind of link for mapping between an *Attribute* node (level 2) and an atomic context *Value* node (level 3) and is denoted by green arc. The green arcs may be timed or untimed. A timed arc is inscribed with the maximum delay counted from the current time instant to obtain the value of the corresponding *Attribute* node. The inscription uses third braces to enclose the time delay. An untimed link has no such annotations. The implications of using timed links are elaborated in section 5.3. All the links to connect three levels are inscribed with corresponding mapping functions.

We can formally define our graph-based context model as 3-tuple $G_c = (N_c, E_c, F_c)$, where

-$N_c$ consists of four types of nodes present in three levels of the model: i) *State* Nodes $N_s$, ii) *Entity* Nodes $N_e$, iii) *Attribute* Nodes $N_a$ and, iv) *Value* Nodes $N_v$. The *Value* nodes again consist of atomic value nodes $N_{v1}$ and composite value nodes $N_{v2}$. Here, $N_v = N_{v1} \bigcup N_{v2}$, $N_c = N_s \bigcup N_e \bigcup N_a \bigcup N_v$, and $N_s \cap N_e \cap N_a \cap N_v = \varphi$. *State* Node $N_s$ is inscribed with information of the form $[P_a, A_a, t_a]$, where $P_a$ and $A_a$ are the set of context parameters and the set of attributes respectively, that indicate a contextual change at time instant $t_a$. An *Entity* Node $N_e$ is inscribed with the name of a context parameter. An *Attribute* node $N_a^j$ ($j > 0$) characterizing an *Entity* node $N_e^i$ ($i > 0$) is inscribed with $N_e^i.N_a^j$, where the dot '.' stands for 'of', i.e., $N_a^j$ is an attribute of context entity $N_e^i$. An atomic context *Value* node $N_{v1}^k$ holds a 2-tuple information $< N_e^i.N_a^j, v_k>$, where the first tuple represents an attribute and the second tuple represents the corresponding value. A composite context *Value* node $N_{v2}$ is the combination (AND or OR) of one or more atomic context values.

- $E_c$ consists of two types of arcs: Relationship arcs $E_R$ and Link arcs $E_L$. Here also, $E_c = E_R \bigcup E_L$ and $E_R \cap E_L = \varphi$. $E_R$ consists of 3 types of relationship arcs: i) $E_{R1}$ to represent the relationships between entities (level 2), ii) $E_{R2}$ to represent the partial dependency relationships among the attributes (level 2), iii) $E_{R3}$ to represent the total dependency relationships among the attributes (level 2). $E_L$ also consists of three types of link arcs: i) Red Arcs between *State* nodes and *Entity* nodes, ii) Blue Arcs between *State* nodes and *Attribute* nodes, iii) Green Arcs between *Attribute* nodes and atomic *Value* nodes.

- $F_c$ represents arc expression functions corresponding to arcs that belong to $E_c$. $E_{R1}$ has the associated arc expression function $F_{R1}^i:(N_e^k \rightarrow N_e^l)$ with $i,k,l > 0$. $F_{R2}$ (corresponding to $E_{R2}$) and $F_{R3}$ (corresponding to $E_{R3}$) are inscribed with the rules of the form '*IF* $C_a$ *THEN* $C_b$', where $C_a$ and $C_b$ are two atomic contexts. The Red Arcs are inscribed with a mapping function $f:(N_s(P_c) \rightarrow N_e)$, where $N_s(P_c)$ is the set of context parameters from a context state **S** at level 1 and $N_e$ is the set of *Entity* nodes at level 2 and $f$ is a one-to-one (but not onto) mapping function between a context parameter and an *Entity* node. The Blue Arcs are inscribed with a mapping function $g:(N_s(A_c) \rightarrow N_a)$, where $N_s(A_c)$ is the set of attributes from the context state **S** at level 1 and $N_a$ is the set of attributes at level 2. In this case also, $g$ is a one-to-one (but not onto) mapping function between each attribute of **S** to each *Attribute* node. Expressions for Green Arcs are basically for value assignments in the variables corresponding to attributes at level 2. The timed Green Arcs are annotated with time delays $d$ and the arc expression will be in the form $\mathrm{Val}(a_i) \leftarrow (v_i, d)$, where $\mathrm{Val}(a_i)$ represents the value of the attribute $a_i$ (variable) and it is expected to be assigned to a value of $v_i$ after time delay $d$. The untimed Green Arcs are expressed by $\mathrm{Val}(a_i) = v_i$, i.e. the attribute variable $a_i$ is assigned to a value of $v_i$ at current time instance. The general structure of our proposed business process context model has been shown in figure 2. Each of the levels in the figure are shown by large rectangular boxes. We can see that inside the level 2 box, three medium sized rectangular boxes are there covering the entire area. The three insider boxes contain a number of *Entity* nodes and the corresponding *Attribute* nodes. These boxes basically contain the lists of contexts from three different perspectives of organization, role and external factors.

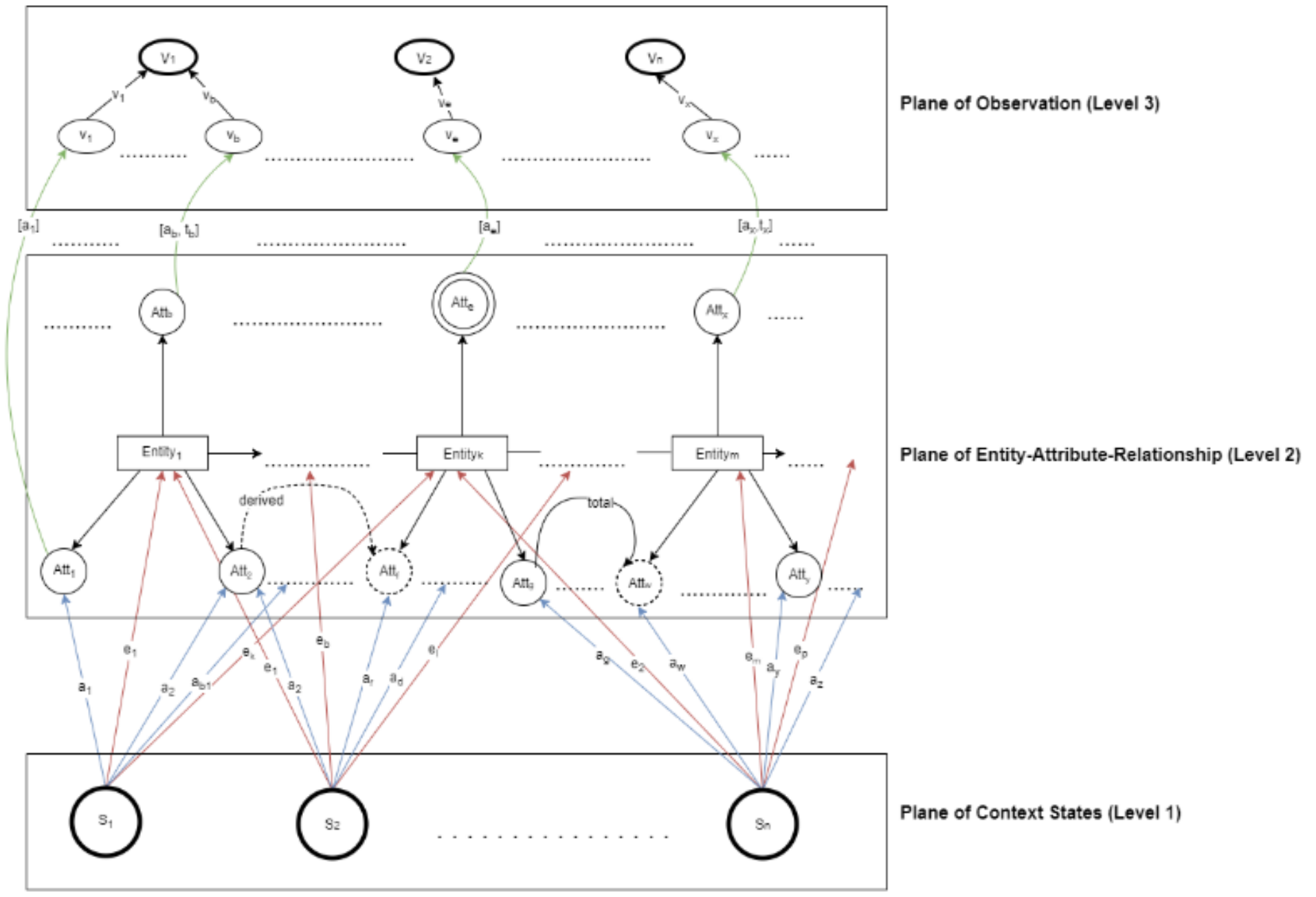

Figure 2: General structure of our proposed business process context model

The business process context model is adapted according to a contextual situation at a particular instant of time. For a business process activity that is affected by the contextual changes, a sub-graph of the adapted model is instantiated at run time on the basis of the context state for that activity. At level 1, there may exist more than one context states. The context states will instantiate corresponding entities, attributes and accordingly the relationships in the context model. There may also be overlapping of nodes and arcs in different instantiated sub-parts. Based on these instantiations, the overall contextual value would be determined and reflected at the *Plane of Observation*.

### 4.4 Reasoning Mechanism for Context Model

Our proposed business process context model supports rich set of context expressions to make the business process understand the semantics of various contexts. The context model offers ease of reasoning as we propose a context predicate for encapsulating the four tuples of a context. The predicates basically characterize the contexts based on their general categories such as organization, role or external factors. The names of the predicates reflect the specialization of the general categories. For example, 'resource' and 'product' are two specializations of the context category 'organization'. Similarly 'patient', 'caregiver' are few specializations of the context category 'role' and, 'season', 'temporal' are some specializations of the category 'external factors'. A sample predicate with four tuples of a context can be represented as "Resource(Network, Connectivity, = , Poor)". Another 'resource' type context can be "Resource(Patient_Bed, Availability_Count, =, 6)". Many instances for a context can be created in a business process environment. Suppose, our example business process uses network connections provided by two different operators 'BSNL' and 'Reliance'. So, there are two instances of context parameter 'Network' namely 'BSNL_Network' and 'Reliance_Network'. For both the instances, the attributes would be same, but the values of attributes may be different. Our proposed context model does not impose any restrictions over the types of attribute values.

A complex context expression can be constructed by performing Boolean operations namely conjunction, disjunction and negation over the context predicates [43].

i) **Conjuction (AND) of Contexts:**

To illustrate the idea behind this, let us take an example: Patient(X, Emergency_Need, =, Patient_Bed) AND Resource(Patient_Bed, Availability_Count, =, 6) refers to the context that the emergency need (need to be hospitalized immediately) of patient X can be fulfilled by a hospital which has six beds available at that time. This kind of 'AND' operation works on the matching arguments within the context predicates under consideration, ignoring the issue whether the context predicates are of same or different categories. When two contexts are from same category, 'AND' operation can be applied when the context parameters or their instances are same for both the contexts. The result of the operation would provide the detailed or required information about the context parameter at a particular time. The syntax used in this case is given below:

$\wedge(\text{Context Predicate})_{\text{Context Parameter}}$ [Contextual Situation (CS)] ………… (1)

We clarify that, for obvious reasons, the contexts present in **CS** are encapsulated by context predicates. When two contexts are from different context categories, ANDing between them is performed only when attribute value of previous context becomes the context parameter (or instance) of the next context. The syntax that is used here is as follows:

$\wedge_{\text{Previous.Attribute_Value = Next.Context Parameter}}$ [Contextual Situation (CS)] ………… (2)

For a different purpose, 'AND' operation can also be applied among context predicates that belong to same context category, when it is necessary to know, for example, what are the available network resources being used by the business process, how many patients are currently being diagnosed etc. In this case, the context predicates are ANDed based on particular conditions. The simple syntax for conditional ANDing of context predicates is as follows:

$\wedge(\text{Context Predicate})_{\text{Condition}}$ [Contextual Situation (CS)] ………… (3)

The condition in the above expression can be on context parameters, attributes and their associated values.

We illustrate expression **(3)** this using a simple example: Suppose the **CS** contains a list of contexts: Resource(BSNL_Network, Connectivity, =, Very Poor), Season(Weather, Status, =, Rainy), Resource(Reliance_Network, Connectivity, =, Average), Healthcare_Assistant(Y, Status, =, Present), Healthcare_Assistant(Y, Working_Experience, =, Good), Caregiver($Z_1$, Expertise, In, Childcare), Caregiver($Z_1$, Status, =, Present), Caregiver($Z_2$, Expertise, In, Arthritis), Caregiver($Z_2$, Status, =, Absent). If we want to know the overall status of network connectivity, we simply follow the syntax:

$\wedge(\text{Resource})_{\text{Context parameter=Instance of 'Network'}}$ [Contextual Situation (**CS**)], which would result in Resource(BSNL_Network, Connectivity, =, Very Poor) AND Resource(Reliance_Network, Connectivity, =, Average).

Also, if we want to know which caregivers are currently available whose expertise is in Artharitis, we follow the syntax:

$\wedge(\text{Caregiver})_{\text{((Context Attribute=Status \&\& Value=Present) AND (Context Attribute=Expertise \&\& Value=Arthritis))}}$ [Contextual Situation (**CS**)], which would result in **NULL**, because no caregivers who have expertise in Arthritis are present at this moment in the healthcare kiosk. If there were more than one caregivers with expertise in Arthritis and were present in the kiosk at that time instant, the list containing all of them would be retrieved.

ii) **Disjunction (OR) of Contexts:**

Let us illustrate this idea using another simple example: Resource(BSNL_Network, Connectivity, = , Very Poor) OR Resource(BSNL_Network, Connectivity, =, No) refers to the context that the network connectivity is either very poor or there is no signal at all. This kind of OR operator is applicable on contexts belonging to same categories. With the same attributes of the contexts under comparison, the OR operation is applied only when either the instances of context parameter are same with differing attribute-values, or attribute values are same with different instances of the context parameter. The example that we have just given applies OR operation on two contexts of 'Resource' categories and in both cases, the instances and attributes are same, but the attribute values are different. If we consider another example to illustrate the other case where OR operator can be applied: Resource(BSNL_Network, Connectivity, = , Very Poor) OR Resource(Reliance_Network, Connectivity, = , Very Poor) is interpreted as the situation when either network (two different instances of context parameter 'Network') suffers from poor connectivity. To be applied, 'OR' operation follows two syntaxes for two cases. For first case when context parameter instances and attributes are same but attribute values are different, the syn tax that is used is:

$$\vee(\text{Context Predicate})_{(\text{Parameter Instance AND Attribute})}\ [\text{Contextual Situation (CS)}] \quad \ldots\ldots\ldots (4)$$

For the second case, when context parameter instances are different, but attribute values are same, the syntax that is used is:

$$\vee(\text{Context Predicate})_{(\text{Attribute AND Attribute_Value})}\ [\text{Contextual Situation (CS)}] \quad \ldots\ldots\ldots (5)$$

iii) **Negation (NOT) of Contexts:**

Consider another example: ¬Patient(X, Suffering, from, Malaria) means that patient X is not suffering from malaria.

Using the combinations of the basic operations, our model supports more complex context expressions. In this case, the operations will be performed on multiple contexts from same or different categories without any conditional restrictions. We also can use the operations like 'Addition' or 'Subtraction' of context values in restricted areas of the business process application domain. Then the operations will be made only between two compatible operators and on their numeric attribute values. Suppose, number of employees in a healthcare centre is currently 10, which is represented by the context Manpower(Healthcare_Assistant, Count, =, 10). Six new employees have joined in the health centre and this situation is represented as Manpower(Healthcare_Assistant, Recruitment, =, 6). The 'addition' operation (10+6=16) can be applied on these two contexts and the resultant context is Manpower(Healthcare_Assistant, Count, =, 16). In this way, our proposed context model can construct and express many complex contexts to reflect the underlying nature of context states or contextual situations in the business process.

## 5. Design of Context-Aware BPMN Model (C-BPMN)

In the proposed work an attempt has been made to offer a means for expressing the context information within existing BPMN. We have modified BPMN in a way such that the process model becomes adaptive in the continuously changing contextual situation. The possible contexts of a business application are captured through our proposed graph based context model. This context model is interpreted with the existing BPMN in an integrated way for expressing the dynamic nature of process model. The generalized BPMN model has been extended with incorporation of additional constructs within BPMN to make it compatible for interpreting the underlying semantics of the effective contextual situations.

### 5.1 Extension on Generalized BPMN model

We propose extension on existing BPMN model of a business process from the upper abstraction view of the model. At the upper abstraction level, we conceptualize the business process model as an exhaustive collection of different sub-goals and each sub-goal is achieved by an individual activity. So, at upper abstraction level, the BPMN model is a sequential collection of activities, with a start and end event. At many lower levels, these activities are decomposed into many tasks, activities, events, gateways etc. The extended BPMN model is termed as "Context-Aware BPMN (C-BPMN)" that consists of all the constructs of generalized BPMN like events, gateways, tasks, activities, arcs etc. in addition with a new construct called 'Contextual Event'. 'Contextual Event' is a special type of event correlated with an activity at the upper abstraction level of the BPMN model. An activity is preceded with a contextual event which corresponds to a context state associated with an activity in the BPMN model. The motivation behind introducing 'Contextual Event' is that a contextual event has only one outgoing edge like other events in the BPMN model and will only direct to one execution pathway depending on the values of context state. Being exposed to all possible pathways depending on different types of contexts using annotated gateways is not an expected and efficient solution to the intended end-users.

From the notational point of view, the special event is represented with a doubly-outlined bold circle to make it different from other existing events and can act as a trigger, which means it reacts to contextual situations using a program module 'catchContext' and can throw a result (in terms of process fragments) using another program module 'throwActivity'. 'catchContext' program module is defined as a set of operations on a set of parameters that are obtained from the Contextual Situation (CS) effecting the whole business process at an instant of time. This module basically catches the real changes in the current state of the contextual event associated with an activity at an upper abstraction level. The programming construct of 'catchContext' module is shown below:

Suppose the latest observed Context State ($\mathbf{S}_a$) of an Activity ($B_a$) is denoted by $[P_a, A_a, t_a]$, where $P_a = \{p_a, p_b \ldots p_g\}$ is the set of context parameters that would affect the execution of $B_a$. On the other hand, $A_a = \{p_i.a_a, p_i.a_b, \ldots p_k.a_k\}$ is the set of attributes that indicate a context change in the activity at time instant $t_a$.

catchContext (**CS**) // The Contextual Situation **CS** at time $t_n$ holds $[P_n, A_n, t_n]$

{

$\mathbf{S}_a := [P_a, A_a, t_a]$; // Latest observed Context State of Activity $B_a$

If ($t_n > t_a$), Then

$\mathbf{S}_c := \mathbf{CS} \sim \mathbf{S}_a$; // The binary operation Difference ('~') computes contextual changes made in $\mathbf{S}_a$ by **CS**

// $\mathbf{S}_c$, a temporary variable, is represented as $[P_c, A_c, t_c]$, where $P_c = P_n \sim P_a$; $A_c = A_n \sim A_a$;

// $t_c = t_n$ if $P_c \neq \varphi$ and/or $A_c \neq \varphi$; $t_c = t_a$, otherwise

If $\mathbf{S}_c \neq \varphi$, Then $\mathbf{S}_a := \mathbf{S}_c$;

Return $\mathbf{S}_a$;

}

A Contextual Event $\mathbf{C}_a$ associated with Activity $B_a$ is annotated with a 2-tuple information $\langle \mathbf{S}_a, V_a \rangle$, where $\mathbf{S}_a$ is the context state of the activity $B_a$ and $V_a$ is the value of the corresponding context state. Based on $\mathbf{S}_a$ in $\mathbf{C}_a$, a sub-part of the proposed context model is instantiated. A *State* node in level 1 of the context model, that corresponds to $\mathbf{S}_a$ is activated and the portions in level 2 and level 3 that are associated to the *State* node are also instantiated accordingly. Finally the value $V_a$ is obtained at $\mathbf{C}_a$ from the observed value at the composite *Value* node at level 3 in the context model. The program module 'throwActivity' associated with $\mathbf{C}_a$ uses the value $V_a$ to derive a process fragment (a set of activities under a business process sub-goal). Process fragments are developed by different service providers or business process providers at pre-design time depending on different sub-goals and different contextual situations and are stored in a process fragment repository. Our primary assumption is that a business process is accompanied with a process fragment repository where all the sub-goals corresponding to the activities in the business process are kept stored for future reference. We also assume that an individual sub-goal may be accomplished by many process fragments in different contextual situations. Under the same sub-goal, an individual process fragment can be designed to cope with one or more contextual situations. But, for the sake of simplicity, we consider, in this paper, that there is one to one correspondence between the composite context values and available process fragments. It may also be possible that contextual situations for two different sub-goals are same, but in this case, the process fragments will obviously be different. So, to search for a fragment in the repository, the sub-goal and value of the composite context must be specified. The block diagram of a process fragment repository is shown in table 1.

Table 1: Internal Constructs of a Generalized Process Fragment Repository

| **Sub-Goal** | **Value(Composite Context)** | **Process Fragment** |
|---|---|---|
| Sub-$\text{Goal}_1$ | Value(Composite_Context $1_a$) = $V_{1a}$ | Process Fragment $k_1$ |
| | Value(Composite_Context $1_b$) = $V_{1b}$ | Process Fragment $k_2$ |
| | …………… | …………………… |
| | Value(Composite_Context $1_m$) = $V_{1m}$ | Process Fragment $k_{\mathrm{m}}$ |
| Sub-$\text{Goal}_2$ | Value(Composite_Context $2_a$) = $V_{2a}$ | Process Fragment $l_1$ |
| | Value(Composite_Context $2_b$) = $V_{2b}$ | Process Fragment $l_2$ |
| | …………….. | ………….. |
| | Value(Composite_Context $2_n$) = $V_{2n}$ | Process Fragment $l_n$ |
| ……….. | …………….. | …………… |
| Sub-$\text{Goal}_N$ | Value(Composite_Context $N_a$) = $V_{Na}$ | Process Fragment $f_1$ |
| | Value(Composite_Context $\mathrm{N_b}$) = $V_{Nb}$ | Process Fragment $f_2$ |
| | ……………… | …………….. |
| | Value(Composite_Context $\mathrm{N_j}$) = $V_{Nj}$ | Process Fragment $f_j$ |

In table 1, all the variables we have used ($N$, $a$, $m$, $n$, $j$, $k$, $l$, $f$) are positive non-zero integers. Retrieving a specific process fragment requires search in the process fragment repository. First, the sub-goal of a particular activity is searched among the set of sub goals listed in the process fragment repository. The list of sub-goals corresponding to activities are indexed using 1 to $N$ ($N > 0$), where $N$ is the number of activities; i.e. sub-goal $K$ ($1 \leq K \leq N$) corresponds to activity $K$ in the BPMN model. So, the search of a sub-goal requires constant time. Once the sub-goal of the activity is matched, the corresponding list of context-values are searched to find the intended context-value of the same activity. If found, the corresponding process-fragment is retrieved. In this paper, we assume that all the fragments are available at design time. But it is also possible that a particular contextual situation is not mentioned in the list. Then, for obvious reason, no process fragment exists for that situation under that particular sub-goal. In this regard, two possibilities are there. The first possibility is that there should be internal structural adaptations by the BPMN model, or the second possibility is that the concerned contextual situation has not been anticipated by the service or business process developers at design time and so there is no information regarding how to handle this kind of situations. Details regarding this issue are discussed in the next section. For example, the sub-goal of an activity "Patient Registration" in our example scenario is the process of registration of a patient by a healthcare employee at the kiosk. The value of the composite context associated with the activity is obtained from our proposed business process context model. Based on the sub-goal and the value of the composite context associated with an activity in the business process model, appropriate process fragment is retrieved from the process fragment repository at run-time. The construct of the program module 'throwActivity' is shown below:

```
throwActivity (SG_a,V_a) // SG_a is the sub-goal of activity B_a and
 {                        // V_a is the value of corresponding composite context
   Go to: SG_a in the indexed Sub-goal list of process fragment repository;
      For j = 1 to k      // k is the total number of context-value entries corresponding to Sub-goal SG_a
         If (V_j == V_a)  Then    //V_j is the j-th composite context value for the Sub-goal SG_a
              Return (V_a, Process_Fragment m_j);   //m_j is the process fragment corresponding to V_a
              Stop();
         Else Return (V_a, NULL);  // no process fragment is selected for activity B_a under the value V_a
         End If
      End For
}
```

If for sub-goal $K$, there are $k1$ number of contextual situations with corresponding $k1$ process fragments, the search performed by the module will take $O(k1)$ time.

### 5.2 Integration of Context Model with Extended BPMN Model

The Context-Aware BPMN model (C-BPMN) is designed through the integration of our proposed context model and the extended BPMN model where the former works above the latter. The Contextual Events serve the purpose of integration between the two models. Two types of links exist between the two models: one type of links for propagation of information about the Context State of a Contextual Event existing in the BPMN model to corresponding state node present at level 1 in the context model, and another type of links for propagation of the value of a corresponding composite *Value* node at level 3 to the same Contextual Event at the BPMN model. Figure 3 shows how a Context Model can be integrated with the Extended BPMN Model. This integrated model is the Context-Aware BPMN (C-BPMN) model.

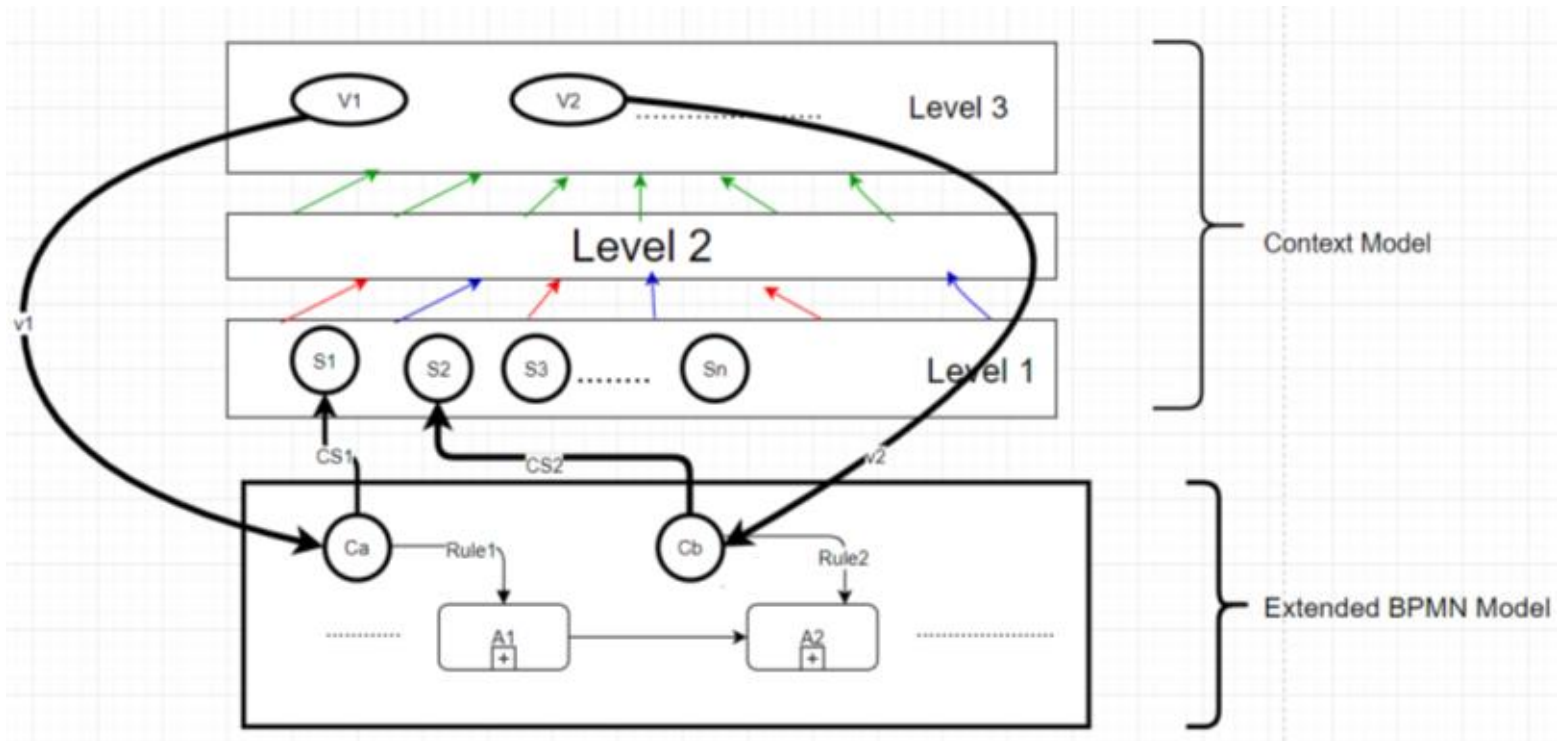

Fig 3: Integration between Context Model and Extended BPMN Model

Our proposed Context-Aware BPMN Model can be formally represented by 5-tuple C-BPMN:
($\mathbf{G}_c$, $\mathbf{EG}_b$, **M**, $\mathbf{Ex}_M$, $\mathbf{I}_S$) where,

1) $\mathbf{G}_c$ is the proposed Context Model.
2) $\mathbf{EG}_b$ is the Extended BPMN model.
3) **M** is the set of links between the two models.
4) $\mathbf{Ex}_M$ is the arc expression function of the links in **M**.
5) $\mathbf{I}_S$ is the ideal state of the integrated model. The 'ideal' state is defined as the state of the integrated model when the BPMN model represents an ideal situation. The ideal situation also is a set of contexts called 'ideal' contexts which are also represented and expressed by the context model.

Suppose $\mathbf{G}_b$ is our generalized BPMN model, where **EVN** is the set of existing events and **ARC** is the set of links defined among existing events, activities and gateways. Now, we define our extended BPMN model with two modifications:

i) $\mathbf{EVN}_m = \mathbf{EVN} \cup \{\mathbf{C}_a, \mathbf{C}_b \ldots \mathbf{C}_n\}$, where $\mathbf{EVN}_m$ is the modified set of events with the addition of new Contextual Events $\mathbf{C}_i$, $i = a, \ldots, n$, where $n$ is the number of activities in the BPMN model
ii) $\mathbf{ARC}_m = \mathbf{ARC} \cup \{\mathbf{a}_a, \mathbf{a}_b \ldots \mathbf{a}_n\}$, where $\mathbf{ARC}_m$ is the modified set of arcs with the addition of new arcs between the contextual events and activities. An arc $\mathbf{a}_i$ basically connects the contextual event $\mathbf{C}_i$ and the activity $\mathbf{A}_i$.

The arc $\mathbf{a}_i$ is inscribed with an adaptation rule $\mathbf{R}_i$ which will define which action (for adaptation) to take on the associated activity after a process fragment is selected under a certain contextual situation. The adaptation rule is in the form of "*IF* (**VAL**== $V_i$) **AND** (**Sel_Frag** == $m_j$) *THEN* (**Action** = $Act_i$)", where the antecedent part of the rule takes two parameters **VAL** (value of the composite context, as observed from the business process context model) and **Sel_Frag** (the process fragment, as returned by 'throwActivity' module) and the consequent part determines an adaptive action that should be taken to cope with dynamically changing contextual situations. We assume that, there exist many adaptation rules for a business process model, keeping in mind different combinations of values of composite contexts and corresponding process fragments. The rules are stored in business process rule repository. Based on the values of composite contexts and the selected process fragments, adaptation rules that are to be applied on the links between contextual events and activities are instantiated at run-time. In the antecedent part of the rule, while the first parameter field **VAL** cannot be empty, the second parameter field **Sel_Frag** may accept NULL values, i.e. no process fragment would be selected.

### 5.3 Adaptation Strategies for Varying Contextual Situations

We visualize each activity in the upper abstraction level of the BPMN model as a graphical node with three compartmentalization such as 'Previous Link', 'Sub-Goal Section' and 'Next Link'. As the upper abstraction level of the BPMN model is a sequential chain of activities, an activity is preceded and followed by two different activities, except those two activities which are associated with start and end events. So, the 'Previous Link' of an intermediate activity holds its preceding activity in the chain and the 'Next Link' holds its next activity in the chain. In case of the activity associated with the start event, the 'Previous Link' would hold 'NULL' and in case of the activity associated with the end event, the 'Next Link' would be 'NULL'. The 'Sub-Goal Section' is basically the data part which holds name of the sub-goal that the activity is going to achieve. The advantage for using such a type of node structure is that our sequential list of nodes can be traversed in both directions whenever necessary. For each activity, there may be five kind of actions in terms of basic adaptation strategies that can be taken based on different contextual situations: "addition of selected process fragment to the current activity", "replacement of current activity", "bypassing the current activity", "changing the execution sequence of a set of activities" and "no change at process level but change at data level". We assume that there will be no violations of policy and consistency in business process model due to these context-aware adaptations.

**i) Addition of selected process fragment to the current activity:**

This action can be taken only when both the parameter fields of a rule are not null. In case of addition of a process fragment to an existing activity $L_1$, the process fragment can be added in two positions: one position is before the activity $L_1$ and another is after the activity $L_1$.

**Case I: Addition before activity $L_1$**

First, we consider an intermediate activity node whose both the links are not null. In this case, the process fragment will be added between the predecessor activity $L_2$ of the activity $L_1$ and the activity $L_1$ itself. Initially, the 'Next Link' part of activity $L_2$ holds $L_1$ and simultaneously, the 'Previous Link' part of activity $L_1$ holds $L_2$. Now, when a process fragment P ($A_1 \rightarrow A_2 \rightarrow A_3$, where $A_1$, $A_2$ and $A_3$ are constituent activities of fragment P) is being added, 'Next Link' part of activity $L_2$ will hold $A_1$ and 'Next Link' part of activity $A_3$ will hold $L_1$; for obvious reason, the 'Previous Link' part of $L_1$ will hold $A_3$ and that for $L_2$ would remain same. One thing to be mentioned here is that, the constituent activities of fragment P are also the activities of upper abstraction level. Now, if activity $L_1$ is associated with the start event, addition of process fragment P before activity $L_1$ would result in following consequences: start event would now point to $A_1$ and the 'Next Link' part of $A_3$ would hold $L_1$; obviously, the 'Previous Link' of $L_1$ would hold $A_3$. On the other hand, if the activity $L_1$ is associated with end event, steps to be followed for addition of process fragment before the activity would be same as an intermediate activity node.

**Case II: Addition after activity $L_1$**

First, we consider an intermediate activity node whose both the links are not null. In this case, the process fragment will be added between the successor activity $L_3$ of the activity $L_1$ and the activity $L_1$ itself. Initially, the 'Next Link' part of activity $L_1$ holds $L_3$ and simultaneously, the 'Previous Link' part of $L_3$ would hold $L_1$. Now, when a process fragment P ($A_1 \rightarrow A_2 \rightarrow A_3$, where $A_1$, $A_2$ and $A_3$ are constituent activities of fragment P) is being added, 'Next Link' part of activity $L_1$ will hold $A_1$ and 'Next Link' part of activity $A_3$ will hold $L_3$; for obvious reason, the 'Previous Link' part of $L_3$ will hold $A_3$ and that of $L_1$ would remain the same. Now, if activity $L_1$ is associated with the start event, steps to be followed for addition of process fragment before the activity would be same as an intermediate activity node. On the other hand, if the activity $L_1$ is associated with end event, addition of process fragment P after activity $L_1$ would result in following consequences: $A_3$ would now point to the end event and the 'Previous Link' part of $A_1$ would hold $L_1$; obviously, the 'Next Link' of $L_1$ would hold $A_1$.

**ii) Replacement of current activity $L_1$:**

This adaptation strategy is followed in three different cases, which are described below:

**Case I: Replacement by selected process fragment**

Suppose, the predecessor activity of an intermediate activity $L_1$ is $L_2$ and the successor activity is $L_3$. Replacement of an activity $L_1$ with a process fragment P ($A_1 \rightarrow A_2 \rightarrow A_3$) means that the 'Next Link' part of $L_2$ will now hold $A_1$, with the 'Previous Link' part remaining same and the 'Next Link' part of activity A3 will now hold the successor $L_3$. So, the 'Previous Link' part of $L_3$ would hold $A_3$ and the 'Next Link' part of $L_3$ will be same as earlier. If the activity $L_1$ is associated with the start event, then replacement of the activity with process fragment P means that start event now points to $A_1$ and the 'Previous Link' part of $L_3$ would hold $A_3$, with the 'Next Link' part of $A_3$ holding $L_3$. On the other hand, if the activity $L_1$ is associated with the end event, the 'Next Link' part of $L_2$ would hold $A_1$ and, $A_3$ would point to the end event. Like the above action, this action can be taken only when both the parameter fields of a rule are not null.

**Case II: Replacement of Role**

In this case, the same activity $L_1$ would be performed by different role. For example, the activity 'Patient Registration' is usually performed by the receptionist. But when the receptionist is absent, the activity can be performed by another healthcare assistant. In this case, both the predecessor and successor activities will be same. In this case, no process fragment is selected for the context value.

**Case III: Replacement of Medium/Mode**

In this case, the same activity $L_1$ would be performed with the help of different medium. For example, the activity 'Bill Payment' is usually performed through net-banking (Payment Mode: Online). But when the network is unavailable, the payment can be made by cash (Payment Mode: Offline). In this case also both the predecessor and successor activities will be same. In this case also, no process fragment is selected for the context value.

Using these three basic replacement policies of strategy (ii), more complex strategies can be developed. For example, when a process fragment has been selected for replacement of an activity $L_1$, the activities under the selected fragment may be performed by the role different from that of activity $L_1$.

**iii) Bypassing the activity $L_1$**

With the same consideration that the predecessor activity of an intermediate activity $L_1$ is $L_2$ and the successor activity is $L_3$, bypassing of activity $L_1$ means that 'Next Link' part of activity $L_2$ will now hold $L_3$ and obviously, the 'Previous Link' part of $L_3$ will hold $L_2$. If $L_1$ is associated with the start event, bypassing means that start event will now point to $L_3$. On the other hand, if $L_1$ is associated with the end event, bypassing means that $L_2$ would now link to the end event. This kind of action can be taken only when no process fragment has been selected for $L_1$ corresponding to a contextual situation.

**iv) Changing the execution sequence of a set of activities**

Sometimes, based on contextual situations, the execution sequence of a set of activities may be changed. This type of action is taken only when no process fragment is selected against a contextual situation. We take a simple example in our example scenario, when a patient is in emergency condition, the activities 'Patient Medical Info Collection' and 'Treatment' are executed before the activity 'Patient Registration'. But in ideal situation, the sequence of activities are: 'Patient Registration' → 'Patient Medical Info Collection' → 'Treatment'. Suppose we have a set of three activities $L_1$, $L_2$ and $L_3$, where $L_1$ is the intermediate activity and $L_2$, $L_3$ are the predecessor and successor activities respectively. So, in ideal situation, the sequence of execution of activities is: $L_2 \rightarrow L_1 \rightarrow L_3$. There may be six different types of execution sequences and for each successful execution of a sequence, many structural adjustments need to be performed internally among the activities. For example, we consider an execution sequence $L_1 \rightarrow L_2 \rightarrow L_3$. In this case, the 'Previous Link' part of $L_1$ will now be replaced by 'Previous Link' part of $L_2$ and the 'Next Link' part of $L_1$ would be $L_2$. The 'Previous Link' part of $L_2$ will be $L_1$ and 'Next Link' part of $L_2$ will be $L_3$. Again, the 'Previous Link' part of $L_3$ will be $L_2$. If $L_1$ is associated with the start event, then $L_2$ will be null. In that case, change of execution sequence will be between $L_1$ and $L_3$. So, the 'Next Link' of $L_1$ would be the 'Next Link' part of $L_3$ and 'Previous Link' of $L_1$ would be $L_3$, with the start event pointing to $L_3$. For obvious reason, the 'Next Link' part of $L_3$ would be $L_1$. On the other hand, if $L_1$ is associated with the end event, change of execution sequence will be between $L_1$ and $L_2$. Then, the 'Previous Link' part of $L_1$ would be replaced by 'Previous Link' part of $L_2$ and 'Next Link' part of $L_1$ will be $L_2$. Hence, the 'Previous Link' part of $L_2$ will be $L_1$ and $L_2$ would point to the end event. In this way, we can also conceptualize the internal structural adaptations for remaining five execution sequences. This adaptation strategy is also used when no process fragment is selected corresponding to a contextual situation.

Let us again consider the same execution sequence 'Patient Registration' → 'Patient Medical Info Collection' → 'Treatment' for an ideal situation. We have seen that the execution sequence of activities changed when a patient was in emergency condition. This kind of contextual situations accept untimed values of composite contexts. But change of execution sequence can also be triggered by timed composite contexts. For the sake of illustration of the effect of timed value of composite contexts on an activity, we can think of a situation from our example scenario where three context changes occur simultaneously. The time is 10.30 am now, the receptionist 'B' who registers a patient is absent at this moment being stuck by road-traffic and is expected to arrive the healthcare centre by half an hour. Healthcare assistant 'Y' is present at 10.30 am at the healthcare centre, but do not know the process of patient registration. The composite context value is [Temporal<Clock, time, =, 10.30 am> AND Receptionist<B, Status, =, Absent> AND Healthcare_Assistant< Y, Knowledge_of_Registration, =, No> AND Healthcare_Assistant<Y, Status, =, Present> AND Environment<Road_traffic, Status, =, Heavy> AND Receptionist<B, Availability, ≤, 11.00 am (30 min)>. This composite context will trigger a change into the execution sequence: the activities 'Patient Medical Info Collection' and 'Treatment' will be executed before the activity 'Patient Registration'. In this case, the assumption is that execution of the activities 'Patient Medical Info Collection' and then 'Treatment' would take time not less than 30 minutes and by this time, the receptionist will be available at the healthcare centre. The contextual situation could also trigger another change in execution like 'Patient Medical Info Collection' → 'Patient Registration' → 'Treatment', if the process for collecting patient's medical information takes at least 30 minutes to execute.

### v) No change at process level but change at data level

There can arise contextual situations for which adaptation cannot be made at the process level but only at the data level. More elaborately, for a contextual situation, no adaptation strategies can be found for addition or replacement of process fragments, bypassing an activity or change in execution sequence, as we have seen in previous strategies. In this case, the contextual situation impacts on the output of an activity, modifying or updating some constituent data sets. Change in output data sets may also be observed at previous four strategies, but changes were also made at process level. As for illustration of the current strategy, we think of a situation that a diabetic patient has come to the healthcare centre for treatment of his fever. But while the symptoms of the patient are being clinically recorded by the attending healthcare assistant, his medical condition suddenly deteriorates heavily. So, that the patient condition is serious should be included in the output of the activity 'Patient Medical Info Collection'. In this case, no adaptation is required at process level but the contextual change should be reflected at output data, so that the adapted business process model can be better-understood by the end-users.

All the five basic adaptation strategies as described can be deeply nested to form more complex adaptation strategies. For example, suppose we are considering the ideal execution sequence $L_2 \rightarrow L_1 \rightarrow L_3$, where $L_2$ and $L_3$ are predecessor and successor activities of $L_1$. We assume, under a contextual situation, the execution sequence should be changed into $L_1 \rightarrow L_2 \rightarrow L_3$. But, under that same contextual situation, $L_1$ should be replaced with process fragment $P_1$ and another process fragment $P_2$ should be added before $L_3$. So, the final activity sequence as part of adaptation would be $P_1 \rightarrow L_2 \rightarrow P_2 \rightarrow L_3$. Another thing that is important to mention here is that all the adaptation strategies are meant for both untimed and timed composite contexts with few restrictions. For a time delay associated with the value of a composite context under a certain contextual situation, addition of process fragment before or after the current activity would be done only after specified time delay. Bypassing an activity $L_1$ for a timed composite context can be interpreted as no need to execute the activity $L_1$ after the specified time delay, with the concerned activity may be executed before that time limit under another contexts. Replacement of an activity $L_1$ for a timed composite context also means that the activity $L_1$ would be replaced by a process fragment after the specified delay and till then the execution of $L_1$ can go on for another contexts. The adaptation strategy for changing the execution sequence for set of activities when the composite context value is timed has been described earlier with our running example. The last adaptation strategy is rather straight-forward with the timed values of composite contexts being replaced by their untimed ones after certain time-limit.

### 5.4 Illustration of proposed C-BPMN model using our running example

Here we illustrate the working mechanism of our proposed C-BPMN model through our example scenario. Suppose, a female patient 'X' of age 45 years has come to the kiosk at 2:00 pm with symptoms of high fever and diabetes. The receptionist 'B' who usually registers a patient has left at that moment and another healthcare assistant 'Z' has been assigned for the role of receptionist. The health condition of the patient suddenly deteriorates while she is waiting at the kiosk and her condition now becomes serious. The caregiver who is available at that time is a child specialist and thus cannot properly handle the patient for treatment. After a preliminary treatment, the caregiver suggests that the patient needs immediate admission for her proper treatment by specialist physicians (when they'll be available). But at that time, no patient beds are available at the kiosk and so, for the patient's admission at near-by hospital, the kiosk makes necessary arrangements like managing an ambulance for transferring patient to the hospital, fixing appointment with specialist physician at the hospital etc. All the arrangements are done over telephone as no internet connection is unavailable at that time due to rain. The medical information of the patient are collected by at kiosk but the gathered information as well as the treatment status (cured/ transferred) cannot be uploaded in cloud server due to unavailability of internet connection. So, without waiting for the internet connection to be available, the process of bill payment at kiosk is done prior to the process of storing information. The bill payment process is usually done online, but as it is not possible at this moment, the billing amount for making necessary arrangements at the hospital by the kiosk is collected by cash, and the business process ends. In this example, we assume that no contextual changes were observed in-between the execution of activities and so, for each activity we note the time instant as '2:00 pm'. So, the contextual situation (**CS**) at 2:00 pm for the concerned business process model at fig 1 is [<Receptionist, Healthcare_Assistant, Patient, Caregiver, Patient_Bed, Weather, Network, Online_Payment>, < Receptionist.Status, Healthcare_Assistant.Status, Patient.Condition, Caregiver.Expertise, Patient_Bed.Availability, Weather.Status, Network.Status, Online_Payment.Status >, <2.00 pm>]. Each of the contexts in the **CS** for the business process does not affect every individual activity. So, we define context state for each activity to capture only the contexts that affect it. The context state $\mathbf{CS_1}$ for activity ‘Patient Registration’ is [<Receptionist, Healthcare_Assistant> <Receptionist.Status, Healthcare_Assistant.Status > <2.00 pm>]. The context state $\mathbf{CS_2}$ for activity ‘Patient Medical Info Collection’ is [<Patient > < Patient.Condition > <2.00 pm>]. The context state $\mathbf{CS_3}$ for activity ‘Treatment’ is [<Caregiver, Patient_Bed > < Caregiver.Expertise, Patient_Bed.Availability > <2.00 pm>]. The context state $\mathbf{CS_4}$ for activity ‘Storage in Cloud’ is [<Weather, Network > <Weather.Status, Network.Status > <2.00 pm>]. The context state $\mathbf{CS_5}$ for the last activity ‘Bill Payment’ is [<Network, Online_Payment> < Network.Status, Online_Payment.Status > <2.00 pm>]. The state nodes $\mathbf{S_1}$, $\mathbf{S_2}$, …, $\mathbf{S_5}$ in the context model are instantiated by the state descriptions $\mathbf{CS_1}$, $\mathbf{CS_2}$,…, $\mathbf{CS_5}$ stored in the contextual events $\mathbf{CE_1}$, $\mathbf{CE_2}$, …, $\mathbf{CE_5}$ respectively. It may also be possible that some state nodes are not triggered at all in a particular situation. In the context model only those entities and attributes are instantiated which are included in the context states. Each instantiated attribute is assigned an atomic value at that time instant and according to definition of each state node, atomic attribute values are combined to get the final composite value for each context state. Each composite value is propagated back to corresponding contextual events. For contextual event $\mathbf{CE_1}$, the composite value $V_1$ [(Receptionist.Status, Absent) AND (Healthcare_Assistant.Status, Present)] and the sub-goal of patient registration trigger to follow the adaptation strategy (ii): replacement of role for execution of the activity by healthcare assistant ‘Z’ instead of the receptionist. For contextual event $\mathbf{CE_2}$, the composite value $V_2$ [(Patient.Condition, Serious)] and the sub-goal of patient’s medical information collection trigger to follow the adaptation strategy (v): no change to be made in the activity under consideration, but that the patient condition has become serious, needs to be reflected in the patient record. For contextual event $\mathbf{CE_3}$, the composite value [(Caregiver.Expertise, Childcare) AND (Patient_Bed.Availability, Not_Available)] and the sub-goal of patient treatment at the kiosk trigger to follow the adaptation strategy (i): addition of the process fragment [‘Appointment Fixing with Specialist Physician at nearby Hospital’ → ‘Arrangement of Ambulance’ → ‘Transfer Patient at Hospital’] after the activity ‘Treatment’ at the kiosk. For contextual event $\mathbf{CE_4}$, the composite value [(Weather.Status, Rainy) AND (Network.Status, Unavailable)] and the sub-goal of uploading the patient’s medical information in cloud server trigger to follow the adaptation strategy (iv): changing the execution sequence between the next two activities namely ‘Storage in Cloud’ and ‘Bill Payment’, i.e. the latter activity would be executed immediately after the ‘Treatment’ activity and the activity ‘Storage in Cloud’ would be the last activity in the chain. For contextual event $\mathbf{CE_5}$, the composite value [(Network.Status, Unavailable) AND (Online_Payment.Status, Not_Possible)] and the sub-goal of bill payment through online trigger to follow the adaptation strategy (ii): replacement of payment mode so that bill can be paid through cash.

With the proposed approach of developing Context-Aware BPMN model, we have overcome the issue of scalability. First of all, only sub-parts of the context model are instantiated at run-time. A Contextual Event associated with each activity in the C-BPMN model react only to the contextual changes that affect the said activity. Depending on the change, the necessary adaptation strategies are taken. Sometimes, the flow of executions among the related activities also changes. Our proposed method ensures that only those parts in a business process model will be affected and modified, that are reactive to context changes at a time instant. So, we can say that scalability issue is not a concern for us.

## 6. Verification and Validation of the Proposed C-BPMN Model Using Colored Petri Net

Like generalized BPMN, our proposed C-BPMN model supports a variety of constructs for good expressiveness and better understandability of the business process in dynamically changing situations. Due to the mix of constructs, the C-BPMN language may prone to a number of semantic errors, including deadlocks, infinite sequences etc. These kind of errors are troublesome during the domain analysis and high level designing of the model, as errors appearing in this time are very difficult and costly to detect [34].

Additional semantic errors can be seen at many times when sub-processes are executed multiple times concurrently or, exception arises while a sub-process is being executed or, when one sub-process calls another sub-process or, when a number of messages are exchanged among different processes and sub-processes. Also, the definition of notation as provided by BPMN is not unambiguous, thereby increasing the probability of more semantic errors existing in the business process model.

At present, the language BPMN lacks a verification technique to detect these kind of semantic errors and needs to be statically analyzed with the help of tools such as Coloured Petri Net (CPN) [34]. CPN is an extension over Petri Net which offers a formal model of concurrent systems. Petri nets are particularly suited to model behavior of systems in terms of "flow", which may be the flow of objects or flow of information [34,15]. Use of colored tokens is an additional feature of CPN with respect to Petri Net, which makes the model more expressive. A CPN model of a system describes the states of the system and the events that can cause the system to change state. By making simulations of the CPN model, it becomes possible to investigate the system behaviour and analyze the properties the model possesses. Using only the BPMN constructs, we would not be able to analyze the behavior of the system whether it is always possible for the system to reach a specified state or whether there is any deadlock in the design or whether the system will provide the service as expected. If any misbehavior or faults of the system are observed after deployment, the rectification will be very problematic. So conversion from a BPMN model to corresponding CPN model is typically done in the early phases of system development for its verification and validation.

The corresponding CPN model of our proposed C-BPMN model is shown in fig 4. The proposed CPN model is a 2-layer hierarchical model with one layer representing the Context model (Layer 1) and another representing the extended BPMN model (Layer 2). In this paper, we only show the conversion of the upper abstraction level of the extended BPMN model into a CPN net called the 'Layer 2' net. In this net, all the BPMN activities (corresponding to different sub-goals) are represented by substitution transitions named '$\text{Activity}_i$', where $1 \leq i \leq n$, with $n$ being the total number of activities at upper abstraction layer. The contextual events associated with each activity are denoted by places named '$\text{ContextualEvent}_i$' with $1 \leq i \leq n$. The start and end events are also denoted by two places inscribed with "Start" and "End" place-names respectively. As per the construction rule of CPN, each transition in this net has one incoming arc from an input place and one outgoing arc to an output place. The concept that an activity in a BPMN model has an input from precedent activity and an output to the subsequent activity, is explicitly visualized in 'Layer 2' net as input and output places associated with an '$\text{Activity}_i$' transition. For each $i^{\text{th}}$ substitution transition, the input place is inscribed with name '$\text{INFO}_{(i-1)}$' and the output place has the name '$\text{INFO}_i$'. This concerned net is decomposed into multiple sub-nets, where each sub-net is dedicated to performing underlying tasks corresponding to each substitution transition and accomplishing necessary adaptations as specified by corresponding contextual event. The sub-nets are constructed using places and transitions with the help of BPMN to CPN conversion rules as stated in [34]. At 'Layer 2', the contextual situation (**CS**) is held by another place with multiple outgoing arcs leading to transitions that are used to represent the modules 'catchContext' associated with the contextual events and accordingly the activities. A '$\text{catchContext}_i$' transition is triggered by effecting contexts obtained from the contextual situation and delivers as output the $i$-th context state directed to the $i$-th contextual event. The context states at the contextual events trigger instantiation of the context model at 'Layer 1' net. The mapping between a '$\text{ContextualEvent}_i$' place at 'Layer 2' net and a '$\text{State}_i$' place at 'Layer 1' net is done using a transition '$\text{PropagateState}_i$'. In the context model, while the *Plane of Context States* consists of a set of places named '$\text{State}_i$' ($1 \leq i \leq n$) to represent the state nodes, the *Plane of Entity-Attribute-Relationship* consists of a number of entities in the form of places named '$\text{Entity}_k$', where ($1 \leq k \leq N$) with $N$ being the total number of contextual entities of a business process for an application and also consists of a number of attributes in terms of places named ‘$A_l$’, where ($1 \leq l \leq M$) with $M$ being the total number of attributes for the business process model under consideration. Every place ‘$Entity_k$’ is associated to some attribute places with the help of a transition ‘$Attributes_k$’. The *Plane of Observation* has $M$ number of places named ‘$value_j$’ with ($1 \leq j \leq M$) to hold atomic value for the attribute ‘$Attributes_j$’. The value-dependency between two attributes is represented by transition named ‘$Dependency_p$’, where ($1 \leq p \leq P$) with $P$ being the total number of dependency relationships among attributes. The combinations of different atomic values of attributes are triggered by transitions named ‘$Composition_i$’ ($1 \leq i \leq n$) and the composite values are obtained at places named ‘$VALUE_i$’ ($1 \leq i \leq n$). The composite values from ‘$VALUE_i$’ places are propagated to the ‘$ContextualEvent_i$’ places at ‘Layer 2’ net. For each ‘$ContextualEvent_i$’ place, the sub-$goal_i$ of activity and the composite value $V_i$ are used to trigger the ‘$throwActivity_i$’ transition. The process_$fragment_i$ that is selected as a result of firing of transition ‘$throwActivity_i$’, is held at the place ‘$Returned_i$’. Based on the value of the place ‘$Returned_i$’, a particular rule $rule_i$ is chosen to make necessary adaptations for the activities. Here all the adaptations as per our strategies are made at the sub-nets and are not shown explicitly in the fig 4, as the result of these adaptations would not lead to any kind of behavioural changes of the model. As our proposed C-BPMN model starts at the start event and ends at the end event, the corresponding CPN model also starts at the place ‘Start’ and ends at the place ‘End’. To check the functional correctness of the model, we place a token at the ‘Start’ place. After simulation, we see that the token is placed at the ‘End’ place, concluding that our model is functionally correct. The behaviour of the model can be captured using the state space analysis.

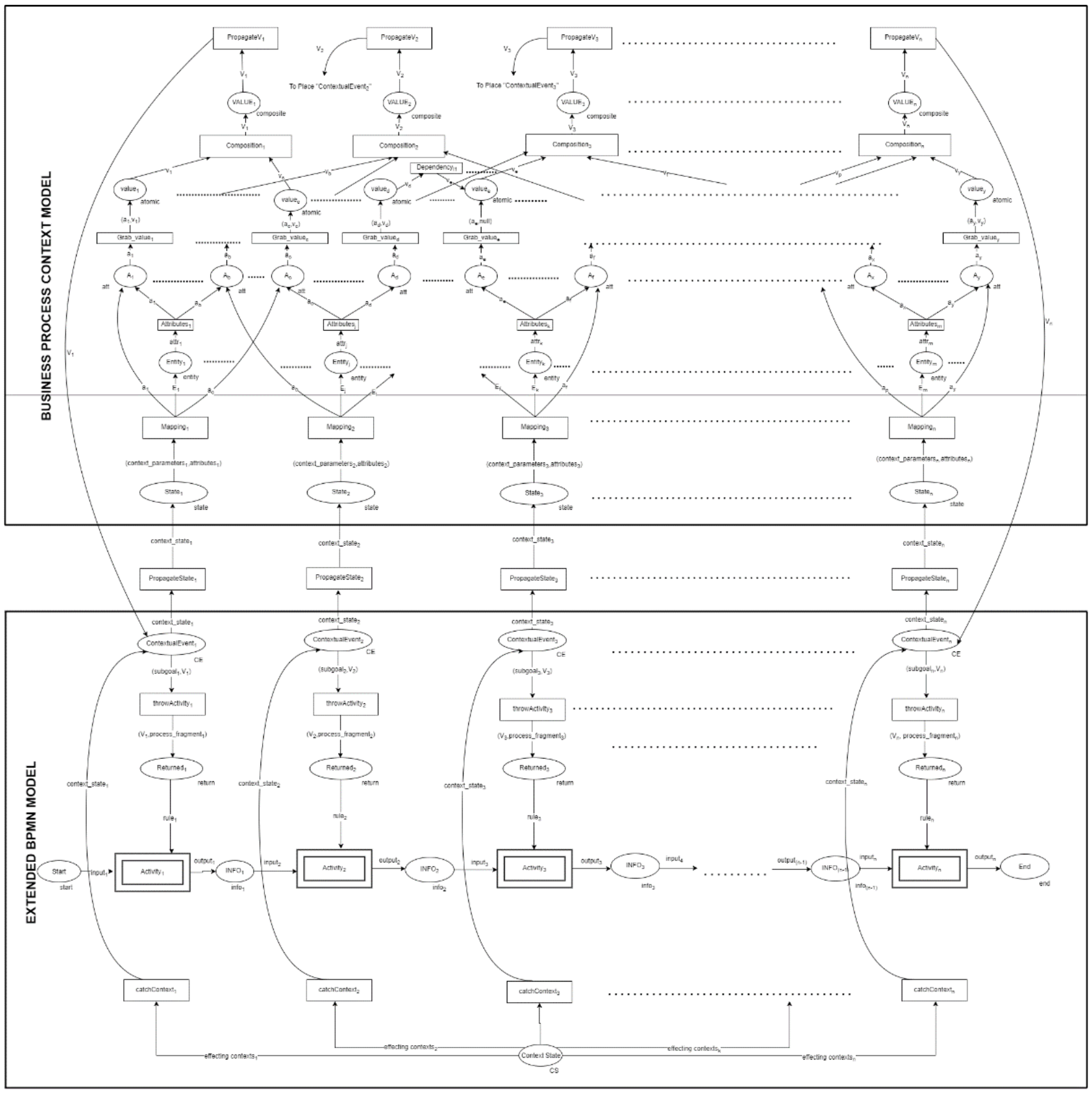

Fig 4: CPN model corresponding to our proposed C-BPMN model

### 6.1 Formal Definition of Proposed CPN Model

Our proposed model can be defined as 9-tuple model G = (Σ, P, T, A, N, C, G, E, M)

-Σ is a finite set of non-empty types, called color sets. These color sets are defined for each place in the model. As our proposed model is a two layer model, the colour sets defined for 'Layer 1' is denoted as $\Sigma_1$= {state, entity, att, atomic, composite}and, that of 'Layer 2' is represented by $\Sigma_2$ = {start, $info_1$, $info_2$, $info_3$, … $info_{(n-1)}$, CS, CE, return}.

-A finite set of places P = $P_1 \bigcup P_2$, where

$P_1$ = {State, Entity, A, value, VALUE}is the set of 'Layer 1' places, where State={$State_i$ | $1\le i \le n$ }, Entity = {$Entity_k$ | $1\le k \le N$}, A = {$A_l$ | $1\le l \le M$}, value = {$value_j$ | $1\le j \le M$} and VALUE = {$VALUE_i$ | $1\le i \le n$} and, $P_2$ = {Start, INFO, ContextualSituation, ContextualEvent, Returned, End} is the set of 'Layer 2' places, where INFO ={$INFO_i$ | $1\le i \le (n\text{-}1)$}, ContextualEvent = {$ContextualEvent_i$ | $1\le i \le n$} and, Returned = {$Returned_i$ | $1\le i \le n$}.

-A finite set of transitions T = $T_1 \bigcup T_2$, where $T_1$ = {Mapping, Attributes, Grab_value, Composition, PropagateV} is the set of 'Layer 1' transitions where Mapping = {$Mapping_i$ |$1\le i \le n$}, Attributes = {$Attributes_k$ | $1\le k \le N$}, Grab_value = {$Grab_value_j$ | $1\le j \le M$}, Composition = {$Composition_i$ | $1\le i \le n$} and PropagateV = {$PropagateV_i$ | $1\le i \le n$} and, $T_2$ = {Activity, catchContext, throwActivity, PropagateState}is the set of 'Layer 2' transitions, where Activity = {$Activity_i$ | $Activity_i$ is a substitution transition with $1\le i \le n$}, catchContext = {$catchContext_i$ | $1\le i \le n$}, throwActivity = {$throwActivity_i$ | $1 \le i \le n$} and, PropagateState = {$PropagateState_i$ | $1\le i \le n$}

A transition that belongs to T would be fired if there exists at least one token at each of the associated input places.

-A is a finite set of arcs such that: P ∩ T = P ∩ A = T ∩ A = Ø.

-N is a node function. It is defined as: A ⊆ (P × T) ⋃ (T × P).

-C is a color function. It is defined from P into Σ such that:

∀ p ⊆ P, there exist one or more elements in Σ such that for every p in P there is at-least one element in Σ.

-G is a guard function. It is defined from T into expressions such that:

$\forall t \subseteq T$: [Type (G(t)) = Bool ∧ Type (Var (G(t))) ⊆ Σ]

In our model, we have not used any guard functions.

-E is an arc expression function. It is defined from A into expressions.

-M is the marking of the proposed model. $M_0$ is the initial marking where the 'Layer 2' places 'Start' and 'Contextual Situation' are marked with tokens.

### 6.2 Implementation of proposed CPN model using CPN tool

Fig 5 gives some glimpses of the implementation of our proposed model. Here, we have modelled a simple scenario of kiosk based remote healthcare solution in India, as described in section 5.4. We have simulated our C-BPMN model for small instances with only one patient, as we are here to check and analyze the behavioural properties of the proposed model. For the implementation purpose, we have incorporated some extra places and transitions in the model, keeping the main essence of the model as shown in fig 4. Also, sometimes the color sets are not same as ones shown in fig 4. If the implemented model shows functional as well as behavioural correctness, the model will be applied to many instances and many patients in real life situations.

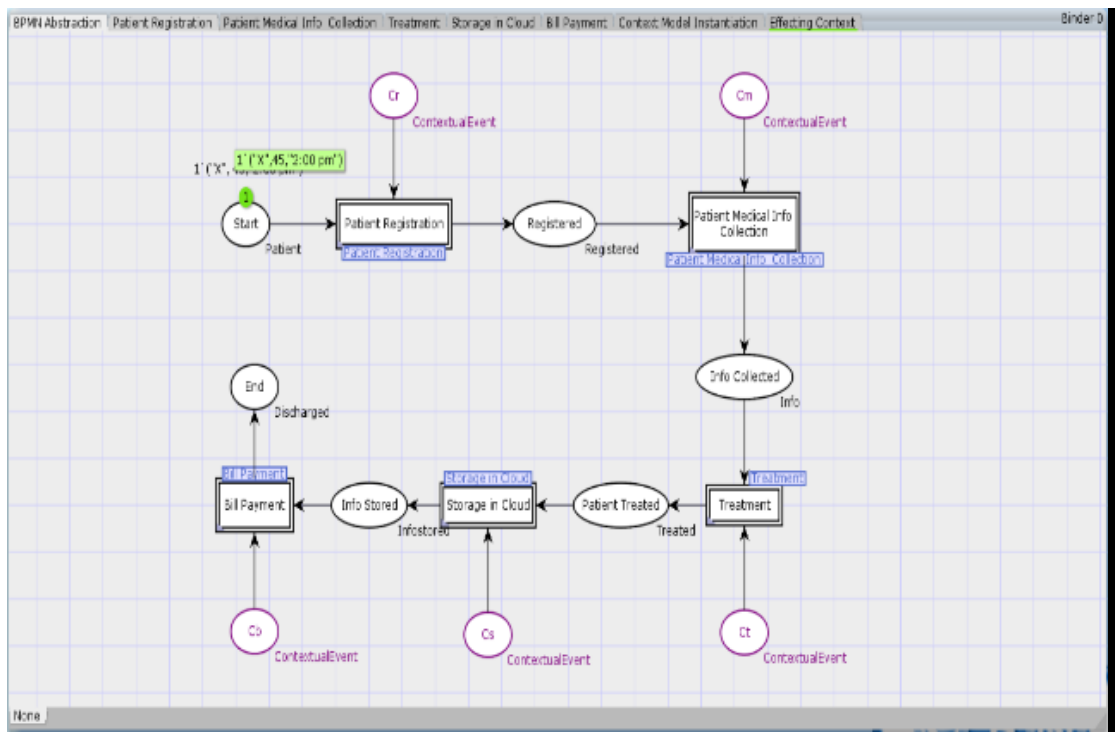

(a)

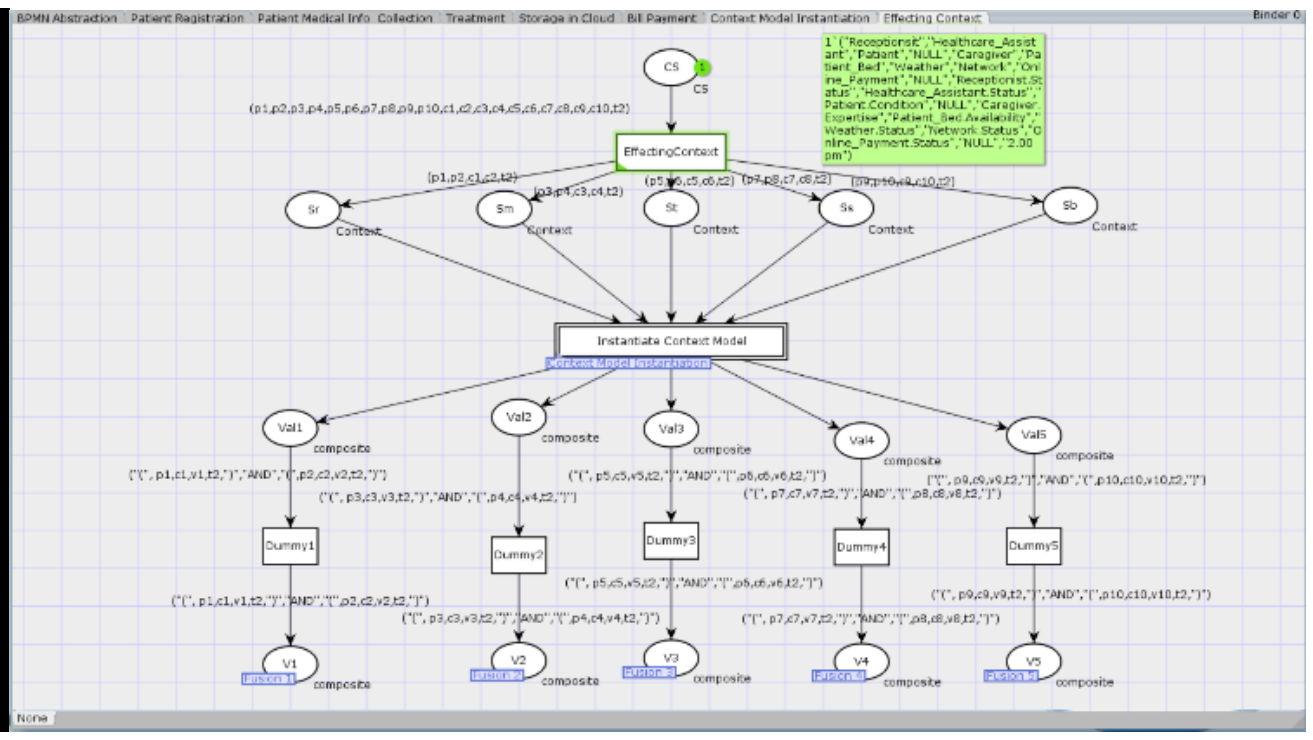

(b)

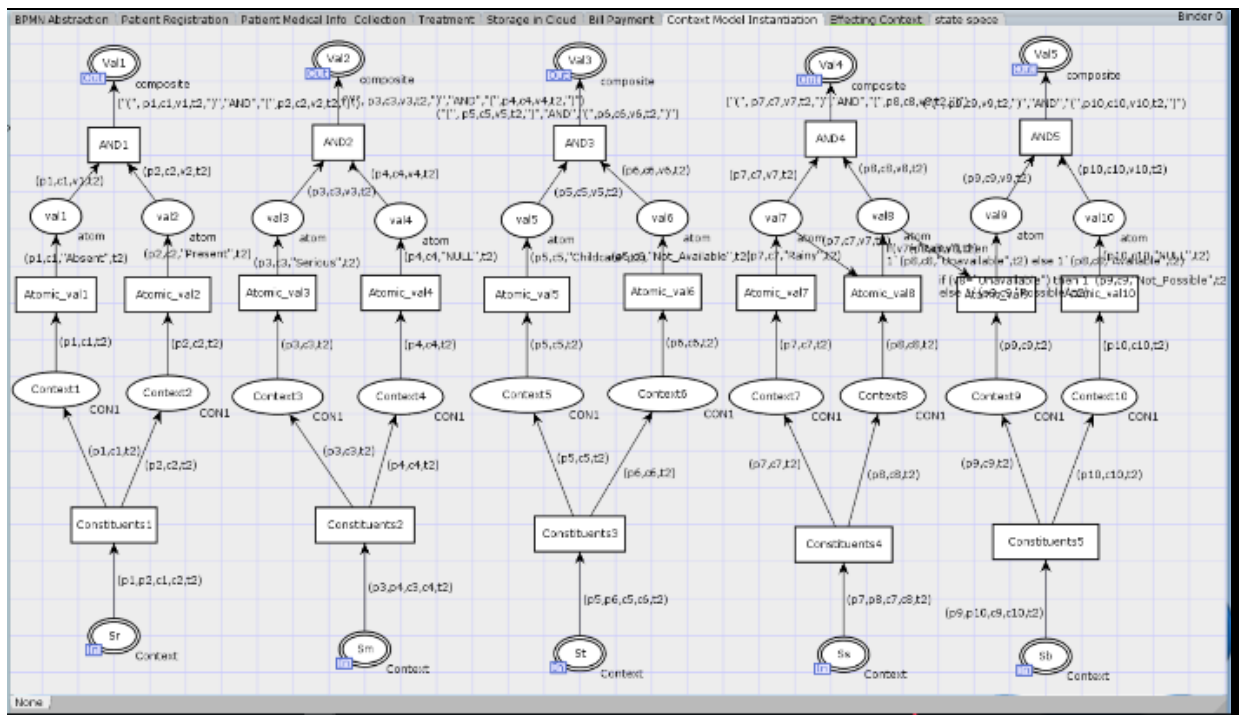

(c)

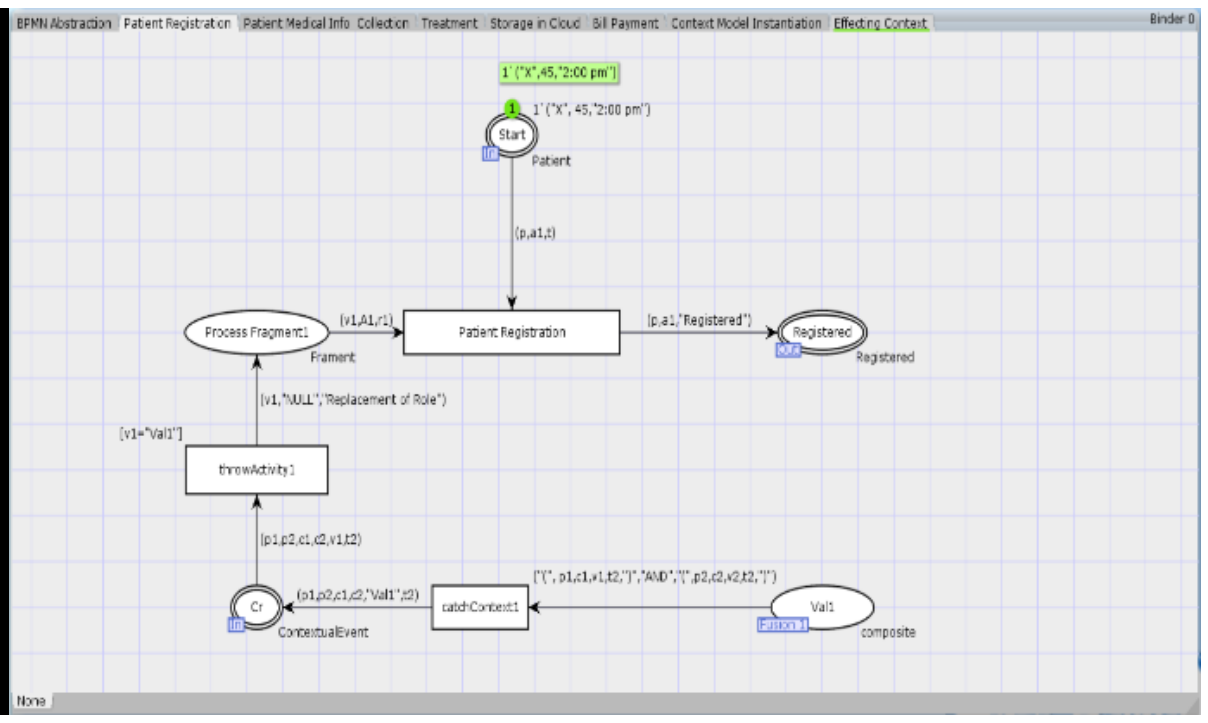

(d)

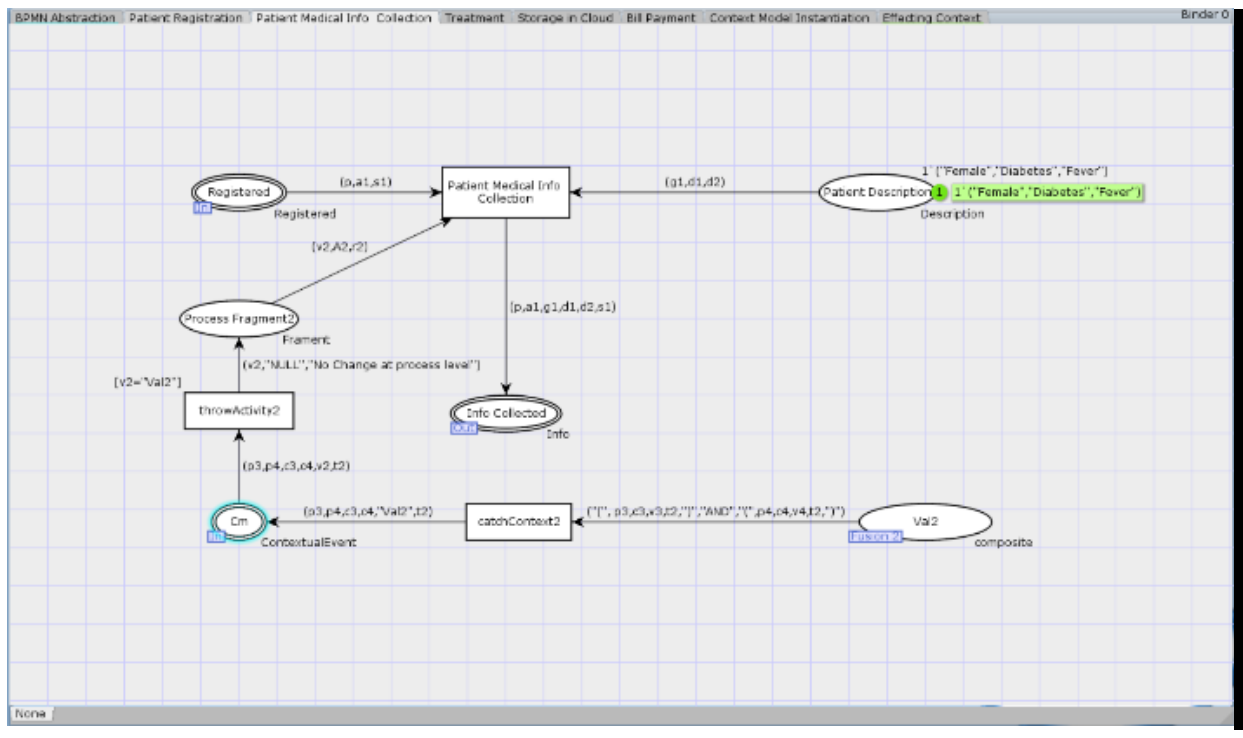

(e)

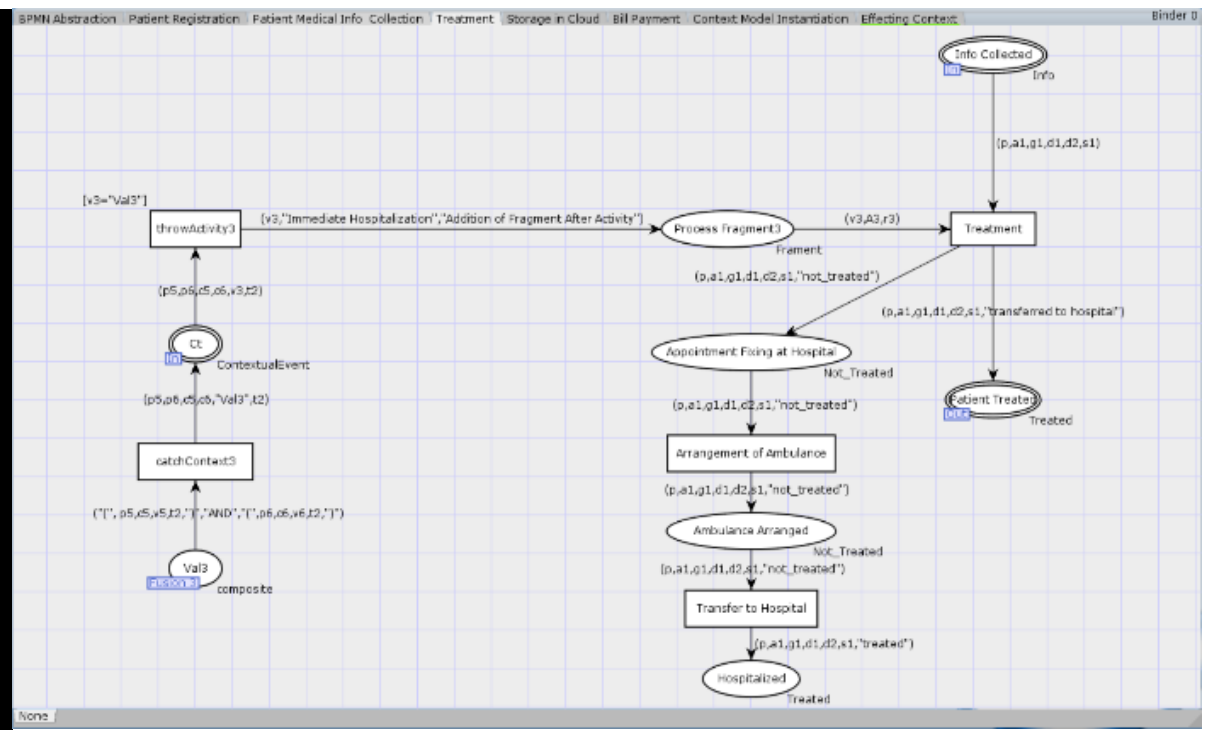

(f)

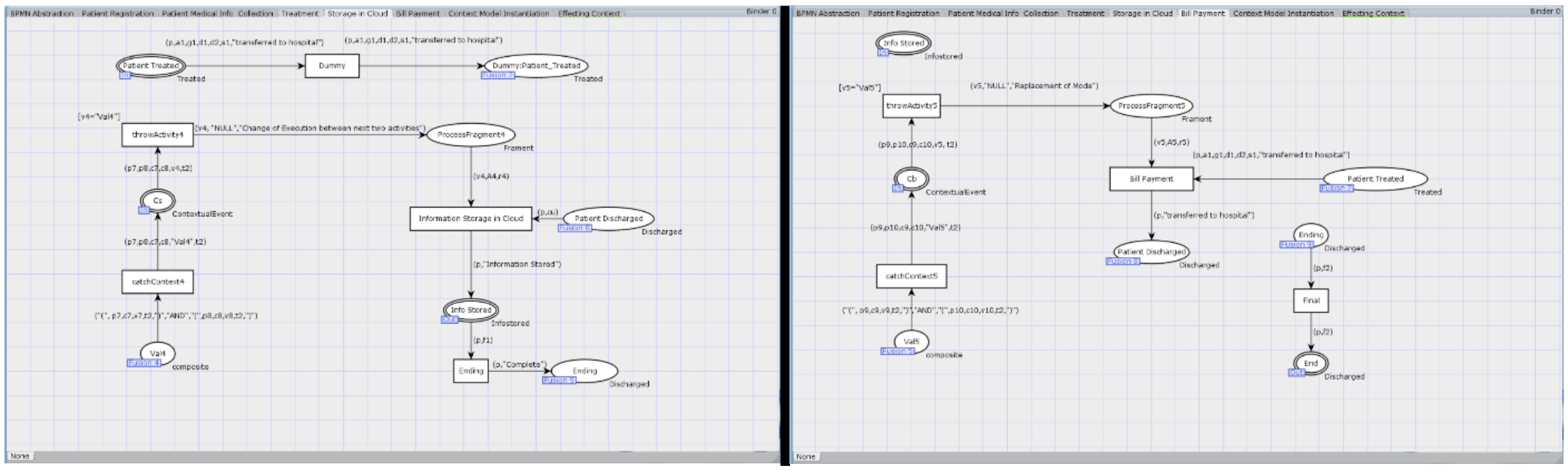

(g) (h)

Fig 5: (a) CPN implementation of top level of 'Layer 2' of C-BPMN model; (b) Selection of context states for contextual events; (c) Instantiation of business process context model; (d) 'Patient Registration' sub-net with necessary adaptations; (e) 'Patient Medical Info Collection' sub-net with necessary adaptations; (f) 'Treatment' sub-net with necessary adaptations; (g) 'Storage in Cloud' sub-net with necessary adaptations; (h) 'Bill Payment' sub-net with necessary adaptations

### 6.3 State Space Analysis of Proposed CPN Model

The validation of CPN model can be done through the state space analysis of the model [36]. The basic idea behind state space analysis is to make a directed graph with a node for each reachable marking and an arc for each occurring binding element [37]. As per the report generated by applying state space tool on our CPN model, the directed graph shows that there are 44575 nodes and 211877 arcs to check reachability from the initial marking of the model to the goal marking. As we can see from fig. 5, the initial marking of the model holds three tokens each at places 'Start', 'Patient Description' and 'CS' and no tokens in remaining places, with the explanation that the model starts execution when a female patient of age 45 years has come to the healthcare centre with symptoms of fever as well as diabetes, and the contextual situation that would affect the execution of the model at that time was noted. The goal-marking corresponds to presence of only one token at the place 'End' with other places being empty. The justification for fixing this marking as goal is that the model would finish execution when the patient will be discharged from the kiosk after executing all the intermediate activities. Remaining places should not hold any token to imply that all the executions of the activities are complete. As we have considered only one patient, when the token has reached the 'End' place, no previous transitions would wait for sequential firing as there are no other patients available. The state space analysis tool concludes that the goal marking is reachable from the initial marking after analyzing the basic behavioural properties [38] like bounded-ness, fairness, liveliness and the home property as described below.

**i)** **Bounded-ness Property**: A place is *K*-safe or *K*-bounded if the number of tokens in that place cannot exceed an integer *K*. A CPN model is said to be bounded if all the places are bounded. Each 'Layer 1' place corresponds to either a state node or an entity node or an attribute node or atomic value node or composite value node. As per our design consideration, for a contextual situation, there would be maximum one token at each place. Each 'Layer 2' place corresponds to either start node or node to hold contextual situation or node to hold activity-output or contextual event or node to hold process fragment returned by 'throwActivity' module or end node. As we are considering only one patient for the sake of simplicity, the start or end node will hold maximum of one token. At any circumstances, there would be only one contextual situation and each contextual event node would also hold maximum of one token at a time. Each node to hold output for an activity will also hold maximum of one token to represent the single output. After calculating the state space, we see that each place has an upper bound of 1 that holds for all reachable markings. This property is important to analyze to determine whether the proposed mechanism is prevented from holding any unnecessary or un-intended information that violates our design consideration. Also it is ensured from this property that no extra tokens are generated or removed during the execution of the model. As per the state space report, our model is 1-safe. So, bounded-ness property holds true for our proposed model. Fig 6 shows results of some standard queries on bounded-ness property related to our model.

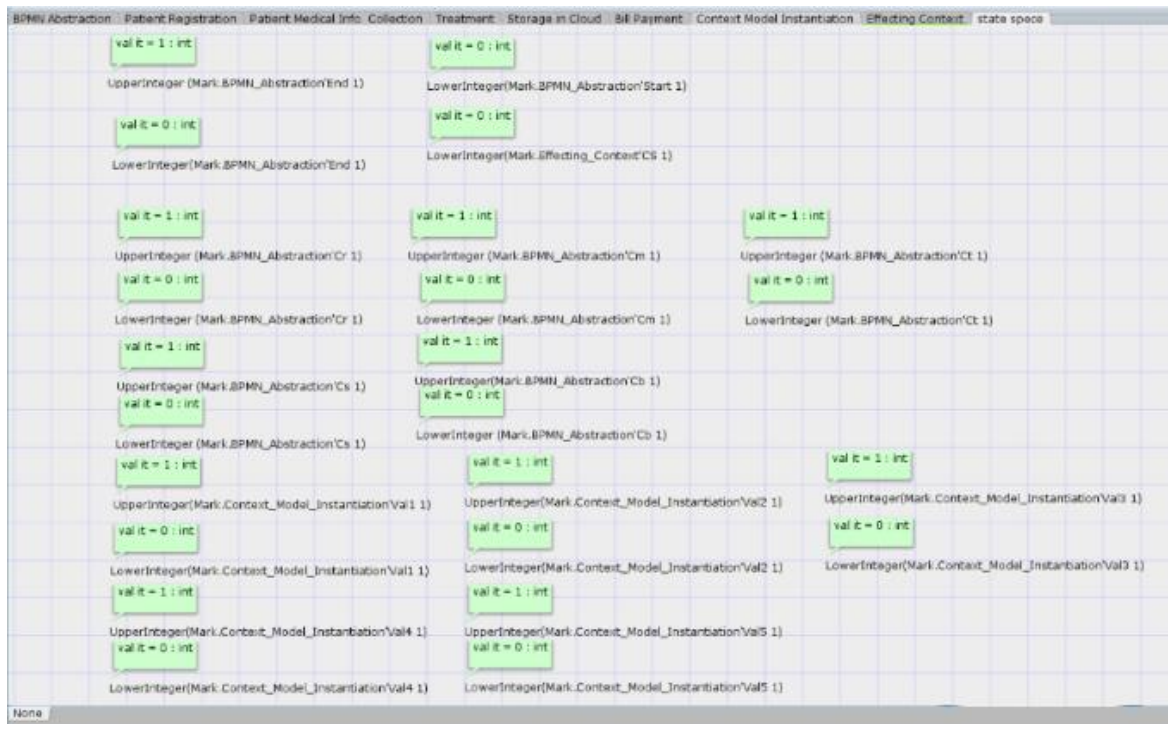

Fig 6: Snapshot of results of some bounded-ness queries related to our model

**ii)** **Live-ness Property:** Liveness is an important property to detect deadlock in the model. A deadlock in the model exists if there is a transition that cannot fire. A CPN is live if each reachable marking if all transitions are fired. More elaborately, there is an occurrence sequence containing all transitions in the model for each reachable marking [39]. As per our design, the context states available in the contextual events trigger necessary instantiations in the context model and all the transitions involved in the instantiated parts of the context model are fired based on firing conditions or rules. So, all the transitions in the proposed model should be live except the goal marking. The state-space analysis report for the CPN model says that there are no dead transition instances in our model and the only dead marking is $M_{44575}$, which means that this marking has no enabling transitions. So, our model supports a deadlock-free design and will not stop execution in any intermediate position without reaching to the place ‘End’. Fig 7 shows results of some standard queries on live-ness property related to our model.

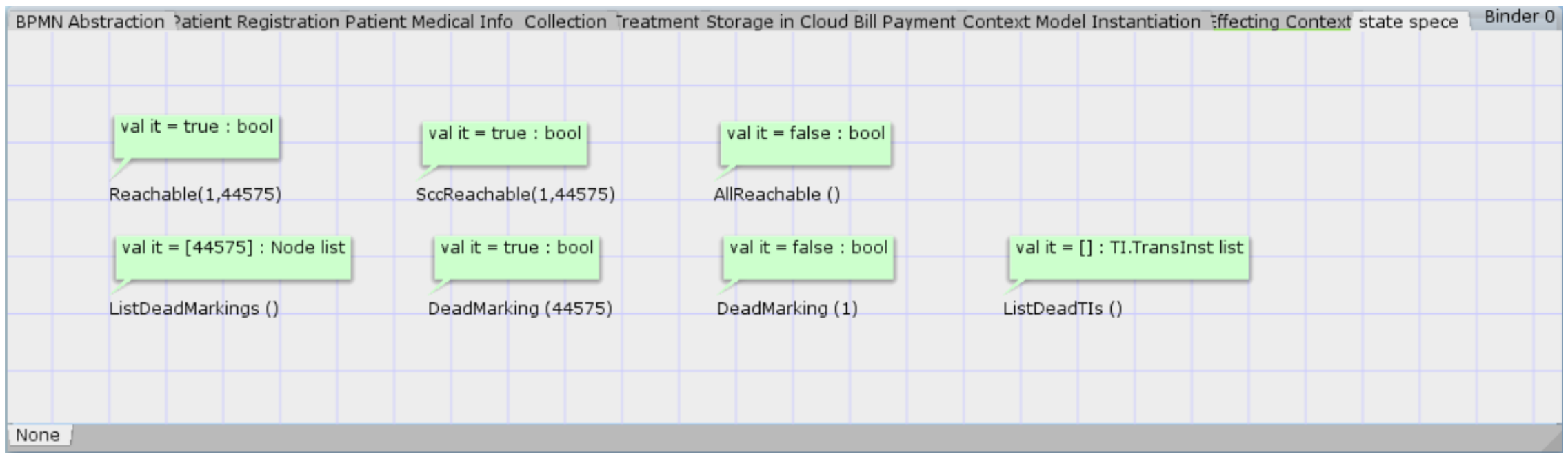

Fig 7: Snapshot of results of some live-ness queries related to our model

**iii) Fairness Property:** Fairness property is used for determining whether a firing sequence in the CPN model is finite or infinite. Infinite occurrence of any transition represents that the execution of the model will not stop after a finite time with finite number of tokens. Infinite occurrence of any transition is a strong drawback after deployment of the model in reality. As per our design, the input token at the place 'Start' would reach to the 'End' place with many intermediate token passing sequences involving many intermediate places and transitions. Apparently, from visual aspects, we see that each place corresponding to a contextual event has an attached loop for passing the description of the context state to the context model ('Layer 1' net) and obtaining the matching context value from the context model. But this kind of loop will not result in infinite looping involving infinite firing sequences of transitions. For the explanation, we say that each place holds only one token and when a transition is associated with two or more input places, all the places should have at least one token to enable or fire the transition. But this scenario when all the input places have tokens will arise only once as per our design. So, the associated transition will be fired only once. The state space analysis report says that there is no infinite occurrence sequence found for the proposed model. So, our design holds fairness property. Fig 8 shows results of some standard queries on fairness property related to our model. A fairness property has four elements {Impartial, Fair, Just, No_Fairness}[40] and all the 'impartial' transitions are subsets of all the 'fair' transitions, which in turn are subsets of all the 'just' transitions. Fig 8 shows that no transitions are unfair in the model.

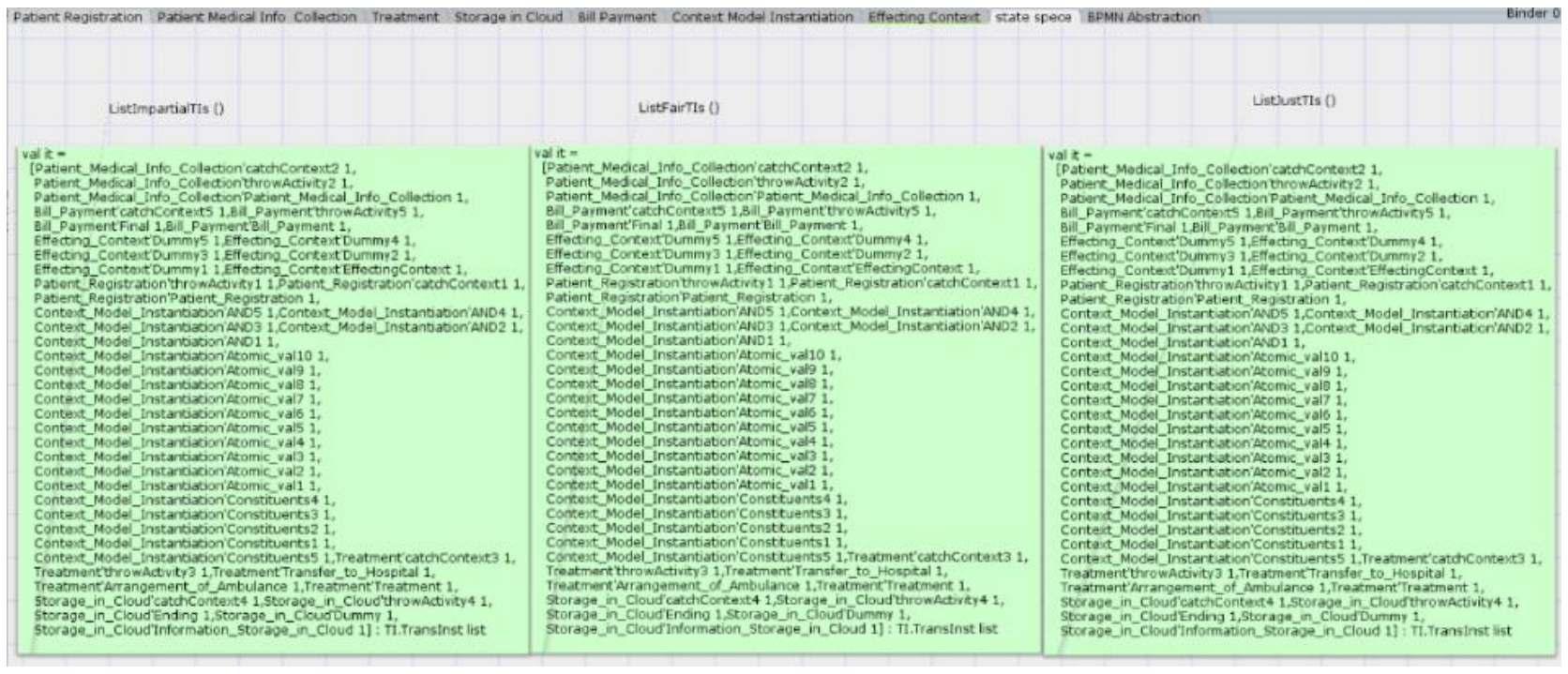

Fig 8: Snapshot of results of some fairness queries related to our model

**iv) Home property:** This property checks whether it is possible to reach the home marking $M_{home}$ from any reachable marking. $M_{home}$ can be any marking. Fig 9 shows results of some standard queries on home property related to our model. As per the report, $M_{44575}$, the goal marking for our model, is reachable from any marking. So, in our case $M_{44575}$ is the home marking.

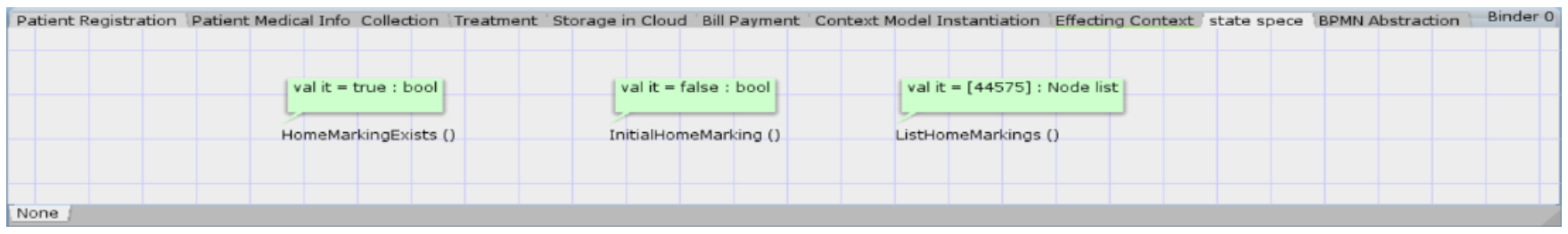

Fig 9: Snapshot of results of some home-property queries related to our model

After properly analyzing all the properties of the CPN model, we claim that the corresponding C-BPMN model also holds the functional correctness as well as behavioural correctness. So, the model will be deployable in any real world situations.

## 7. Brief Discussion on Performance Measures of Proposed Model

Performance analysis is often a central issue in the design, development and configuration of systems. It is not always enough to know that systems work properly, it should also be ensured that the systems work efficiently. Few performance measures for our proposed model is given below:

i) **Execution Time of Proposed Model:** This measure is to calculate the execution time of activities from the start event to the end event in the C-BPMN model. Suppose, each activity in the upper abstraction layer of extended BPMN model takes $\boldsymbol{t_a}$ time to execute all the intermediate tasks using many events (not contextual events) and gateways. As per our assumption, there are $n$ activities in the sequence and each activity is associated with a contextual event. While the 'catchContext' module takes constant time $C_{ct}$, the process of obtaining value for the associated composite context takes $\boldsymbol{t_{cm}}$ time which is basically $\mathbf{O}(N_c^I + E_c^I)$, where $N_c^I$ and $E_c^I$ are the number of nodes and edges in the context model, that are instantiated dynamically according to contexts. The 'throwActivity' module takes $\boldsymbol{t_{th}}$ time, that is basically $\mathbf{O}(K1)$, with $K1$ being the number of process fragments under a sub-goal for different contextual scenarios. We assume that the execution of a selected process fragment takes $\boldsymbol{t_p}$ execution time. So, the total execution time for the extended model is $n$*[$\boldsymbol{t_a} + \boldsymbol{t_p} + \boldsymbol{t_{cm}} + \boldsymbol{t_{th}} + C_{ct}$], which is better than execution time of an alternate model when many possibilities are checked through gateways corresponding to different contextual situations.

ii) **Need of Extra Storage for the Proposed Model:**

a) **Number of additional activities in a business process:** This metric is the extension of the metric NOA [21] and counts the number of extra activities that are required in the extended BPMN process, in addition to the number of existing activities in the generalized BPMN version. As per our design consideration, we have incorporated no extra activities in the extended BPMN model in the ideal situation. When there is a need for structural change in the model due to adaptation to changing environment, only the required process fragments are added to those activities that are affected by the contextual change. So, in that case, the design is considered to be better than the design with consideration of all alternatives using gateways.

b) **Number of extra activities and control flow elements in a business process:** This metric is the extension of the metric NOAC [21] and counts the number of extra activities that take process control flow nodes (gateways) into account. In our design, no extra gateways are required to make our proposed business process model adaptable to dynamically changing situations.

c) **Number of extra control paths through a business process:** This metric is the extension of the metric McCabe's cyclomatic complexity (MCC) [21,41] and is defined as ($E$` - $N$` + 2), where $n$` is the number of extra constructs in the model and $e$` represents the number of extra sequence flow between elements in the model, with $N$` being the total number of $n$` constructs and $E$` being the total number of $e$` sequence flows. We have introduced one new construct called contextual event. So, in our model, $N$` is $n$ and there are also $n$ number of control transfers ($E$`) among the contextual events and associated activities, in addition to the ones in the generalized BPMN model. So, the value of the metric is only 2, which obviously indicates a good design.

d) **Control Flow Complexity (CFC) metric:** This is a standard metric [41] based on MCC to take into consideration the mental state of the designer while modeling a business process. The metric counts the total number of branches in the model including the total number of states arising in the model due to the use of split gateways. This count for our C-BPMN model is same as that of generalized BPMN model, with no additional complexity inducing in the model in-spite of incorporation of adaptability in the model. Using a number of gateways in a BPMN model to cope with contextual situations is not a good alternative modeling approach as the complexity would increase heavily.

e) **Halstead – based process complexity metrics (HPC):** HPC takes data complexity of a business process model into consideration. The primitive measures of HPC are based on number of unique flow elements ($n1$), number of unique data objects ($n2$), total number of $n1$ elements ($N1$) and total number of $n2$ elements ($N2$). Suppose there are $n1$` number of unique flow elements in the generalized BPMN model. As one unique construct has been introduced in the C-BPMN model, $n1 = n1$`+1. Let us consider that there are $n2$` number of unique data objects in the generalized BPMN model. We assume that one new data object associated with a contextual event is introduced in the extended BPMN model to hold data from the 'catchContext' module, the composite context value as well as data from the 'throwActivity' module. So, $n2 = n2$`+ 1. So, $N1 = n + N1$`, with $N1$` being the total number of $n1$` elements in the generalized BPMN model and $N2 = n + N2$`, with $N2$` being the total number of $n2$` elements in the generalized BPMN model. So, Halstead-based process length metric for C-BPMN model is ($n1 * \log_2(n1) + n2 * \log_2(n2)$). Halstead-based process volume metric for the proposed model is $(N1+N2) * \log_2(n1+n2)$ and Halstead-based process difficulty metric for the model is $(n1/2)*(N2/n2)$.

The additional storage of $\mathbf{O}(N_c^I + E_c^I)$ is required at each instantiation of the proposed context model based on different context states, where $N_c^I$ and $E_c^I$ are the number of nodes and edges in each instantiated part. As the above metrics show storage efficiency for the C-BPMN model, this little additional storage does not result in any performance degradation of the model.

## 8. Conclusion

Incorporating context-awareness in BPMN model is an enhancement of the modeling technique in terms of process flexibility and adaptability. There are some works on context awareness in business process models, but no such work has been found on how to integrate context model explicitly into the existing BPMN to achieve the context aware business process modeling. Also, the adaptation strategies are specified in the paper. The designer with domain expertise can design the business process context model and integrates the context model with the extended BPMN model independently. The approach increases the simplicity and understandability of the given approach. This separation also increases the maintainability in terms of addition of new constructs within a business process. The Context-aware BPMN model has also been simulated and validated using CPN tool for analyzing the structural and behavioral properties of the model. The proposed one is proved to have the functional as well as behavioral correctness after doing necessary state-space analysis.